\documentclass[fleqn,usenatbib]{mnras}

\usepackage{amsmath,amssymb,gensymb}
\usepackage{multicol,tabularx}
\usepackage{graphicx,xcolor,xspace}
\usepackage{booktabs,multirow,subcaption,threeparttable}

\usepackage[T1]{fontenc}
\usepackage{ae,aecompl}
\usepackage{textcomp}
\usepackage[capitalize]{cleveref}

\newcommand{\subtext}[2]{\ensuremath{#1_{\text{#2}}}} 
\newcommand{\pccm}{pc\,cm$^{-3}$\xspace} 
\newcommand{\imagebox}[2]{\vtop to #1{\null\hbox{#2}\vfill}} 
\title[SPARKESX Data Challenge]{SPARKESX: Single-dish PARKES data sets for finding the uneXpected --- A data challenge}

\author[S.~Yong et al.]{
Suk Yee Yong$^{1}$\thanks{E-mail: \href{mailto:sukyee.yong@csiro.au}{sukyee.yong@csiro.au}},
George Hobbs$^{1}$,
Minh T.~Huynh$^{2,3}$,
Vivien Rolland$^{4}$,
Lars Petersson$^{5}$,
\newauthor
Ray P.~Norris$^{1,6}$,
Shi Dai$^{6}$,
Rui Luo$^{1}$,
Andrew Zic$^{1,7}$
\\
$^{1}$CSIRO Space and Astronomy, PO Box 76, Epping, NSW 1710, Australia\\
$^{2}$CSIRO Space and Astronomy, PO Box 1130, Bentley WA 6102, Australia\\
$^{3}$International Centre for Radio Astronomy Research (ICRAR), M468, The University of Western Australia, 35 Stirling Highway,\\Crawley, WA 6009, Australia\\
$^{4}$CSIRO Agriculture and Food, Clunies Ross St, Acton, ACT 2601, Australia\\
$^{5}$CSIRO Data61, Clunies Ross St, Acton, ACT 2601, Australia\\
$^{6}$School of Science, Western Sydney University, Locked Bag 1797, Penrith, NSW 2751, Australia\\
$^{7}$School of Mathematical and Physical Sciences, and Research Centre in Astronomy, Astrophysics and Astrophotonics,\\Macquarie University, NSW 2109, Australia\\
}%

\date{Accepted XXX. Received YYY; in original form ZZZ}
\pubyear{2022}

\begin{document}
\label{firstpage}
\pagerange{\pageref{firstpage}--\pageref{lastpage}}
\maketitle

\begin{abstract}
New classes of astronomical objects are often discovered serendipitously. The enormous data volumes produced by recent high-time resolution, radio-telescope surveys imply that efficient algorithms are required for a discovery. Such algorithms are usually tuned to detect specific, known sources. Existing data sets therefore likely contain unknown astronomical sources, which will remain undetected unless algorithms are developed that can detect a more diverse range of signals. We present the Single-dish PARKES data challenge for finding the uneXpected (SPARKESX), a compilation of real and simulated high-time resolution observations. SPARKESX comprises three mock surveys from the Parkes ``Murriyang'' radio telescope. A broad selection of simulated and injected expected signals (such as pulsars, fast radio bursts), poorly characterised signals (plausible flare star signatures) and unknown unknowns are generated for each survey. The goal of this challenge is to aid in the development of new algorithms that can detect a wide-range of source types. We show how successful a typical pipeline based on the standard pulsar search software, \textsc{PRESTO}, is at finding the injected signals. The dataset is publicly available at \url{https://doi.org/10.25919/fd4f-0g20}.
\end{abstract}


\maketitle

\section{Introduction} \label{sec:intro}

The quest to find the unknown objects in our vast universe is one of the most intriguing challenges in modern astrophysics. Astronomical discoveries can be separated into two broad classes, the ``known unknowns'' and the ``unknown unknowns'' \citep{Norris:2017}. The known unknowns relate to discovering more of a particular class of astronomical object with unknown origin, such as fast radio bursts (FRBs) or pulsars. The ``unknown unknowns'' relate to unexpected signals that, if proven to be astronomical in origin, will lead to new areas of astronomical research.

Many astronomical objects and phenomena are discovered using radio telescopes, making radio astronomy surveys ideal for searching for even more astronomical objects \citep{Ekers:2009}. One of the largest steerable single dish radio telescopes is the Parkes ``Murriyang'' telescope. With its large 64\,metre aperture dish and high sensitivity, the Parkes telescope is capable of mapping large areas of the sky and has made many major scientific discoveries in the time-domain astronomy, including FRBs \citep[e.g.,][]{Lorimer+:2007}, pulsars \citep[e.g.,][]{Manchester+:2001}, double pulsars \citep[e.g.,][]{Burgay+:2003}, and rotating radio transients \citep{McLaughlin+:2006}. The observations from these surveys are publicly available from the Commonwealth Scientific and Industrial Research Organisation (CSIRO) Data Access Portal (DAP)\footnote{\url{https://data.csiro.au/collections/domain/atnf/search/}}. The telescope is also involved in the Breakthrough Listen project, a 10\,year-long initiative to search for extraterrestrial intelligence \citep[SETI;][]{Merali:2015,Isaacson+:2017,Price+:2018}. The data from the Breakthrough Listen project are also open access\footnote{\url{http://seti.berkeley.edu/opendata}}.

Unknown signals (specifically SETI) have long been sought with interest and at the same time, with considerable skepticism. Many theories and speculations have been proposed, yet the nature of such sources remains a mystery \citep[see reviews by e.g.,][]{Ekers+:2002,Wright:2022}. One pioneering suggestion was to look for ``beacons'' or indications of technosignatures in the radio frequency range between 1--10\,GHz across interstellar space \citep{Cocconi+Morrison:1959}. Typical SETI experiments search for narrowband spectral signals in frequency-time images or spectrograms which have a Doppler drift rate caused by the relative radial acceleration from the transmitter to the receiver \citep{Drake:1961}. However, such signatures can be similar to terrestrial and/or space-borne radio frequency interference (RFI), which results in a vast number of false candidates  \citep[e.g.,][]{Tarter:2001,Siemion+:2013,Enriquez+:2017,Wright+:2018,Price+:2020,Perez+:2020,Sheikh+:2020,WlodarczykSroka+:2020,Margot+:2021,Gajjar+:2021,Traas+:2021}. A narrowband technosignature verification framework has been established in order to eliminate false events that are attributed to RFI \citep{Sheikh+:2021}.

Instead of directing high power into a narrow frequency window, it is also possible that extraterrestrial beacons are compressed in time having the form of a short broadband pulse \citep{Cole+Ekers:1979}. Radio pulses are dispersed in the interstellar medium and exhibit positive dispersion measure (DM). An extraterrestrial civilisation therefore might deliberately send an artificially induced broadband signal with negative DM \citep{Siemion+:2010,vonKorff:2010,Harp+:2018,Li+:2020,Gajjar+:2021}. Searches for such artificial dispersed signals have been conducted, but none have yet been found \citep{Gajjar+:2021}.

Numerous software packages have been developed to search for individual, dispersed pulse events. Two of the most common open source pulsar search software packages are the PulsaR Exploration and Search TOolkit\footnote{\url{https://github.com/scottransom/presto}} \citep[\textsc{PRESTO};][]{Ransom:2001,Ransom+:2002} and \textsc{SIGPROC}\footnote{\url{http://sigproc.sourceforge.net/}} \citep{Lorimer:2011}. Both have been rigorously tested and maintained, and have been used to discover many pulsed radio sources. With the high growth rate of data volume, there is an increasing demand for real-time or near real-time processing. The next generation telescope surveys will generate petabytes of data per year, triggering the need to handle the data in an efficient way. The advancement of technology has allowed this to become feasible through the innovation of graphics processing units (GPUs), which significantly increase the processing power and speed. Several GPU-based single pulse search software packages have been developed including \textsc{Heimdall}\footnote{\url{https://sourceforge.net/projects/heimdall-astro/}} \citep{Barsdell:2012}, \textsc{peasoup}\footnote{\url{https://github.com/ewanbarr/peasoup}} \citep{Barr:2014,Barr:2020}, and \textsc{AstroAccelerate}\footnote{\url{https://github.com/AstroAccelerateOrg/astro-accelerate}} \citep{Adamek+Armour:2020}. Typically, these algorithms de-disperse the data set at many DM trials before searching for pulsed events in the resulting time series. This can be computationally intensive and various fast de-dispersion techniques have been proposed, such as the fast DM transform \citep{Zackay+Ofek:2017} and the Fourier-domain de-dispersion \citep{Bassa+:2022}. These algorithms assume the dispersion law and hence are not designed to detect other signatures that may be present in the data stream. Several factors including the presence of RFI and poorly matched filters, will affect the sensitivity and completeness of the search \citep[see e.g.,][]{Barsdell+:2012,Keane+Petroff:2015,vanHeerden+:2017}.

Recent developments in artificial intelligence and machine learning have also brought forth new possibilities. Machine learning and deep learning approaches have already been applied to classify single pulses \citep{Zhang+:2018,Connor+:2018,Agarwal+:2020b,Kunkel+:2021}. In addition, numerous data challenges are being setup in the astronomy community \citep{Hlozek:2019}. A few examples are data challenges relating to the Large Synoptic Survey Telescope \citep{Kessler+:2019}, the Australian Square Kilometre Array Pathfinder and Evolutionary Map of the Universe \citep{Hopkins+:2015}, and the Square Kilometre Array \citep{Bonaldi+Braun:2018}.

The aim of this paper is to provide a collection of time-series dataset consisting of simulated and injected signals that are both likely within our data streams (such as FRBs) and unexpected, more generic signals. The artificial signals are inserted into simulated data and also into actual Parkes observations. This work presents a standardised set of tests that have been carefully designed to explore the pros and cons of different algorithms. The baseline results from a standard method are provided for comparison with existing or new pipelines. The data is publicly available and we expect will be used by numerous groups to develop and test their search algorithms prior to applying them to actual observations.

The outline of the paper is as follows. \Cref{sec:data} describes the dataset and the different types of signals generated.  Baseline results using \textsc{PRESTO} are presented in \cref{sec:searchpresto}. In \cref{sec:discussion}, we discuss the limitations of the \textsc{PRESTO} pipeline and challenges in the data. Finally, the conclusion and future direction are provided in \cref{sec:conclusion}. We describe how to download the data challenge in \cref{appsec:sparkesxinfo}.

\section{Data and Simulation} \label{sec:data}

\begin{table*}
  \centering
  \caption{List of simulated system and observation parameters.}
  \label{tab:typedata}
  \begin{tabular}{l *{7}{c}}
  \toprule
  Survey & Time & Sampling & Bandwidth & Central & \# Channel & \# Bit & Reference \\
  & duration [s] & time [$\mu$s] & [MHz] & frequency [MHz] & & & \\
  \midrule
  1-beam of multibeam & 1024.000 & 250 & 288 & 1374 & 96 & 1 & \citet{Manchester+:2001} \\
  13-beam of multibeam & 8400.896 & 1000 & 288 & 1374 & 96 & 1 & \citet{Manchester+:2006} \\
  PAF & 933.888 & 57 & 500 & 1300 & 2048 & 2 & Dunning et al, in prep \\
  \bottomrule
  \end{tabular}
\end{table*}

Our mock ``Single-dish PARKES radio telescope for finding the uneXpected (SPARKESX)'' dataset was created using the \textsc{simulateSearch}\footnote{\url{https://bitbucket.csiro.au/scm/psrsoft/simulatesearch.git}} software package \citep{Luo+:2022}. \cite{Luo+:2022} provides general details about the simulation software, which can simulate a range of source types for different observing modes at various telescopes. Here we explicitly apply that software to produce fake observations from the Parkes radio telescope. The majority of the actual survey data from the Parkes telescope was obtained using a 13-beam, 20\,cm receiver (the ``multibeam''). We simulated three mock surveys. The first mock survey has similar properties to the multibeam survey datasets that make up the bulk of the Parkes data archive, but, in this case, we only simulate data corresponding to a single beam. We do this as many data processing algorithms treat the beams independently. However, in particular for transient events, multibeam information is invaluable for distinguishing between astronomical sources and radio frequency interference. The second mock survey is based on longer-duration observations of the Magellanic Clouds \citep{Manchester+:2006} and here we simulate data for all 13 beams of the multibeam receiver. The third provides an example of the next generation of high time-resolution surveys that will shortly begin with a cryogenically-cooled phased array feed (PAF). The PAF will provide up to 76 beams, but again here we only simulate a single beam. The system parameters are listed in \cref{tab:typedata}. In all cases, we only simulate one polarisation channel.

The simulated data are stored in \texttt{PSRFITS} format\footnote{\url{https://www.atnf.csiro.au/research/pulsar/psrfits_definition/Psrfits.html}} \citep{Hotan+:2004}, which is adapted from the Flexible Image Transport System \citep[\texttt{FITS};][]{Wells+:1981} file format. The data are partitioned into blocks of a specified number of time samples (NSBLK) and are recorded in successive rows (or sub-integrations) in the file. The multibeam survey data are 1-bit quantised with a level setting procedure applied that operates with a rolling time constant of 0.9\,second \citep[see][]{Manchester+:2001}. The first block for the simulated PAF data streams are used to set the 2-bit quantisation and then the levels are held fixed for the remainder of the observation.

For a given set of input events (described below), a total of 50 search mode data files were generated for surveys with a single beam, i.e., the single-beam representations of the multibeam and PAF surveys. For the 13-beam data sets, we only provide injected simulated signals into real data (see \cref{ssec:injectintoreal}).

Our work here is based around developing and comparing algorithms that can identify rare events. We therefore explicitly ensure that the events (1) do not overlap with each other and (2) only form a small fraction of each data file. We also require the means to determine whether an algorithm has failed or succeeded in identifying any event. Each frame, for the 1-beam simulations, either contains one event or no events. In order to achieve this, we reset the 1-bit level setting process at the frame edges. For very bright events, simulated near the edge of the frame this can lead to unrealistic step-changes in the noise properties at the frame boundaries. This is not an issue for the 2-bit data streams.  

The 50 data files for the single-beam mock surveys provide ample data for testing different search algorithms. Each file contains 1000 frames with 10 percent of the data containing a signal of interest (100 events). An overview of the number of generated files is in \cref{tab:sparkesxfile} in \cref{appsec:sparkesxinfo}. In this case, the multibeam nature of the observations are simulated and we do not divide the data set into specific frames and so these simulations are most like the actual observations and can be used to test streaming-based (i.e., not specific to individual frames) search methods.

We do not simulate realistic survey observation times as we simply need the data to be long enough for our tests. Each survey has a different sampling time (see \cref{tab:typedata}). The number of samples per sub-integration (NSBLK) is set to 4096. We have used frame lengths of $\sim 1$\,second corresponding to 1 sub-integration for the single-beam and 4 sub-integrations for the PAF simulations. 

All the simulated data files contain fake radiometer noise. This is modelled \citep[see Section 3.2 of][]{Luo+:2022} assuming a Gaussian distribution with a specific amplitude derived from an assumed system temperature of 21\,K and telescope gain of 0.735\,K\,Jy$^{-1}$ \citep[e.g.,][]{Manchester+:2001,Keith+:2010}. Events that are dispersed in the interstellar medium are modelled using the arrival time difference, $\Delta t$, between two pulses at frequency, $\nu$:
\begin{equation}
\begin{split}
&\text{DM}\ [{\rm pc\,cm}^{-3}]= \\
&\quad\frac{\Delta t}{4.149 \times 10^{-3}\,{\rm s}} \left[\left(\frac{\subtext{\nu}{low}}{1000\,{\rm MHz}}\right)^{\alpha} - \left(\frac{\subtext{\nu}{high}}{1000\,{\rm MHz}}\right)^{\alpha} \right]^{-1},
\label{eqn:dm}
\end{split}
\end{equation}
where $\alpha$ is the spectral exponent. For FRBs and pulsars, $\alpha$ is $-2$. As our events do not overlap and must fit within a single frame, we limited the maximum DM simulated to be 1087\,\pccm for the single-beam from multibeam and 458\,\pccm for the PAF survey.

\subsection{Type of Simulated Events}

For each of the 1-beam data sets, we have simulated the six categories of injected source types that are listed in \cref{tab:simsignalgroup}. The first (\texttt{noise}) has no injected signal and simply contains radiometer noise. The second (\texttt{rfi}) contains radiometer noise and RFI, but no astronomical signal. We then simulate idealised FRB-like signals (\texttt{simplepulse}) that are broadband, have no frequency-dependent structure and exactly follow the frequency-squared dispersion law. The fourth (\texttt{known+rfi}) contains more realistic signals including likely signals from flare stars and FRB-like events that contain frequency-dependent pulse structure. The fifth set (\texttt{unknown+rfi}) of simulations contain more generic source shapes (representing the ``unknown unknowns'') and finally we include all event types (\texttt{combo+rfi}).

\begin{figure*}
\captionsetup[subfigure]{aboveskip=-3pt,belowskip=-2pt}
\centering
\begin{subfigure}[t]{\textwidth}
  \includegraphics[width=\textwidth]{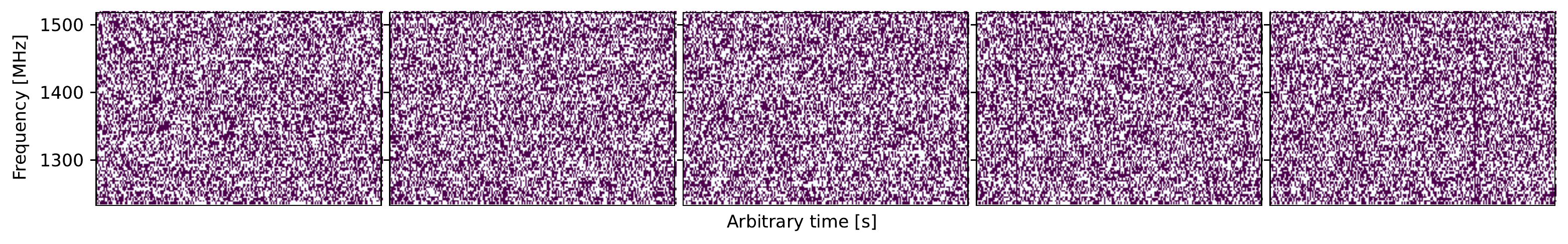}
  \caption{RFI}
  \label{fig:rfi_multi_01}
\end{subfigure}
\begin{subfigure}[t]{\textwidth}
  \includegraphics[width=\textwidth]{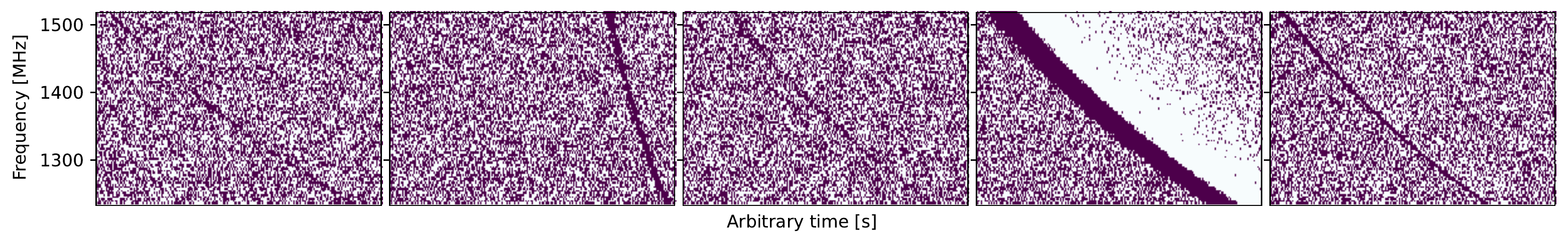}
  \caption{Simple pulses}
  \label{fig:simplepulse_multi_01}
\end{subfigure}
\begin{subfigure}[t]{\textwidth}
  \includegraphics[width=\textwidth]{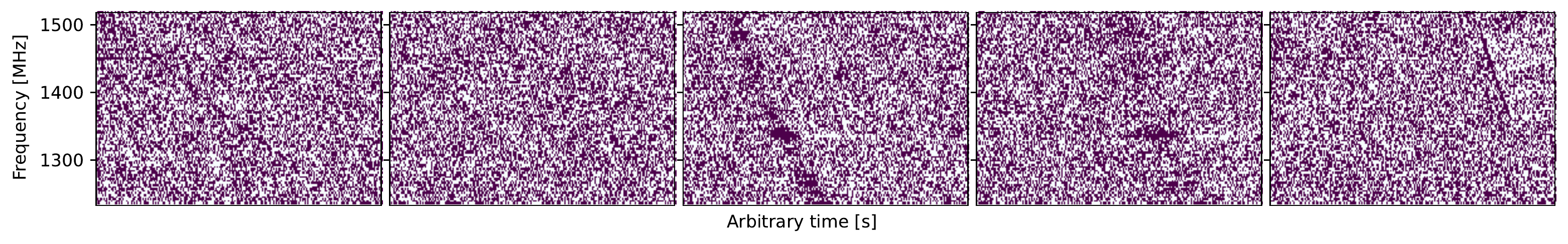}
  \caption{Real pulses}
  \label{fig:realpulse+rfi_multi_01}
\end{subfigure}
\begin{subfigure}[t]{\textwidth}
  \includegraphics[width=\textwidth]{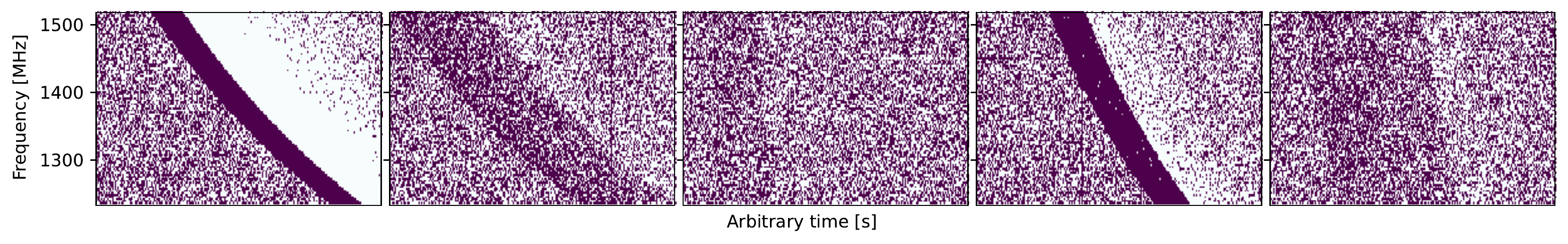}
  \caption{Long pulses}
  \label{fig:longpulse+rfi_multi_01}
\end{subfigure}
\begin{subfigure}[t]{\textwidth}
  \includegraphics[width=\textwidth]{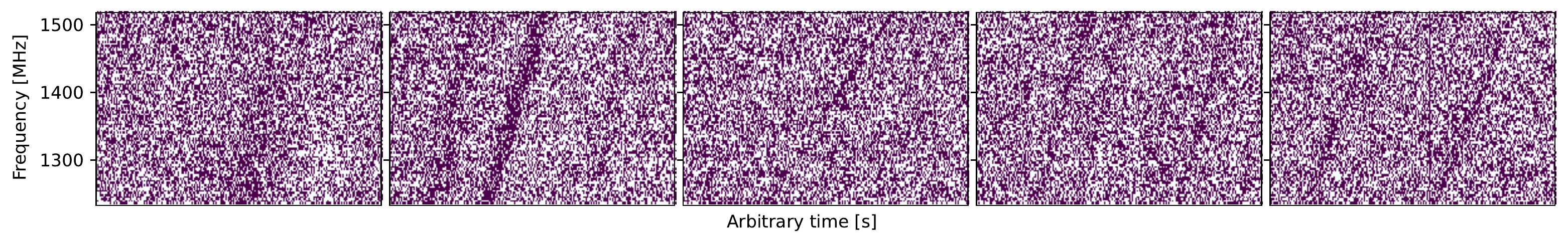}
  \caption{Stellar flares}
  \label{fig:flare+rfi_multi_01}
\end{subfigure}
\begin{subfigure}[t]{\textwidth}
  \includegraphics[width=\textwidth]{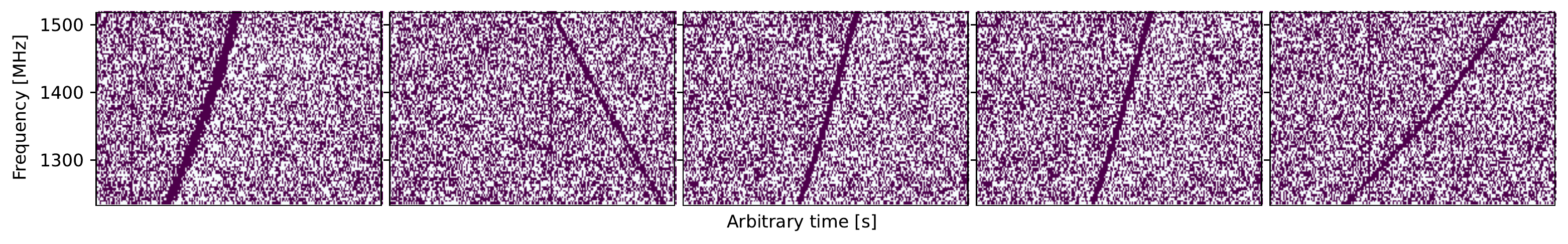}
  \caption{Negative DM pulses}
  \label{fig:ndmpulse+rfi_multi_01}
\end{subfigure}
\begin{subfigure}[t]{\textwidth}
  \includegraphics[width=\textwidth]{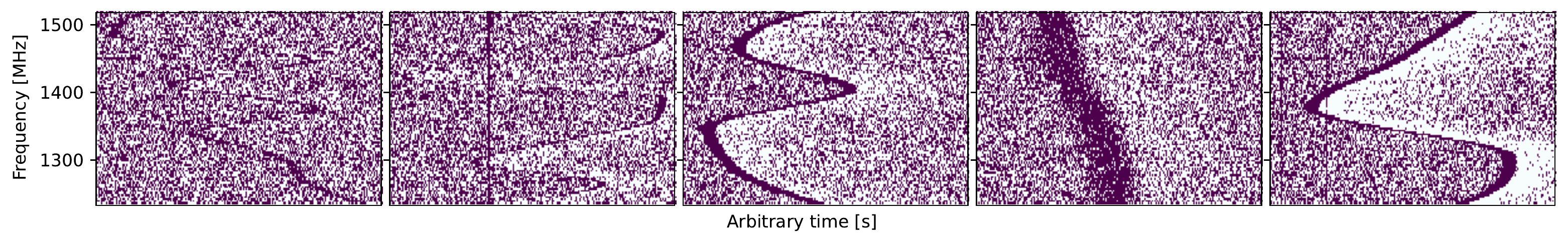}
  \caption{Splines}
  \label{fig:spline+rfi_multi_01}
\end{subfigure}
\begin{subfigure}[t]{\textwidth}
  \includegraphics[width=\textwidth]{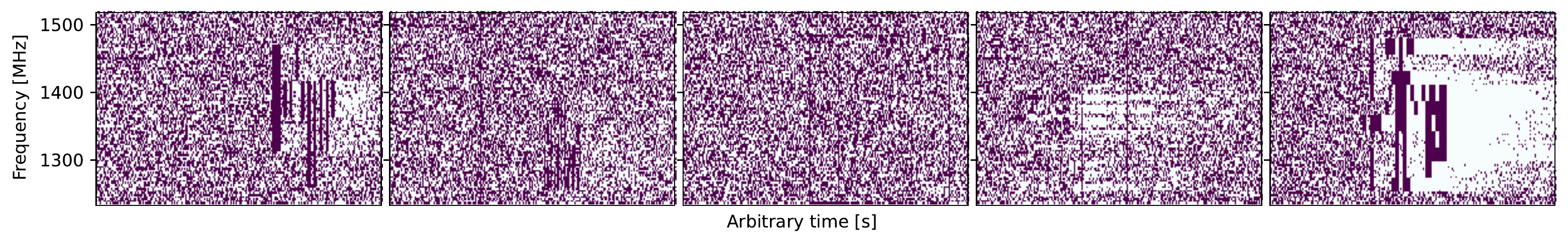}
  \caption{Steganography images}
  \label{fig:image+rfi_multi_01}
\end{subfigure}
\caption{Examples of simulated events for the single-beam from multibeam survey.}
\label{fig:simsignal_multi_01}
\end{figure*}

\begin{figure*}
\captionsetup[subfigure]{aboveskip=-3pt,belowskip=-2pt}
\centering
\begin{subfigure}[t]{\textwidth}
  \includegraphics[width=\textwidth]{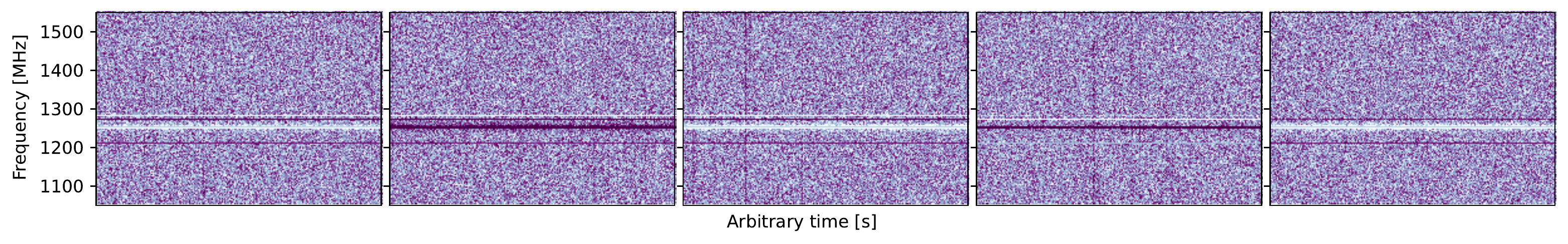}
  \caption{RFI}
  \label{fig:rfi_paf_01}
\end{subfigure}
\begin{subfigure}[t]{\textwidth}
  \includegraphics[width=\textwidth]{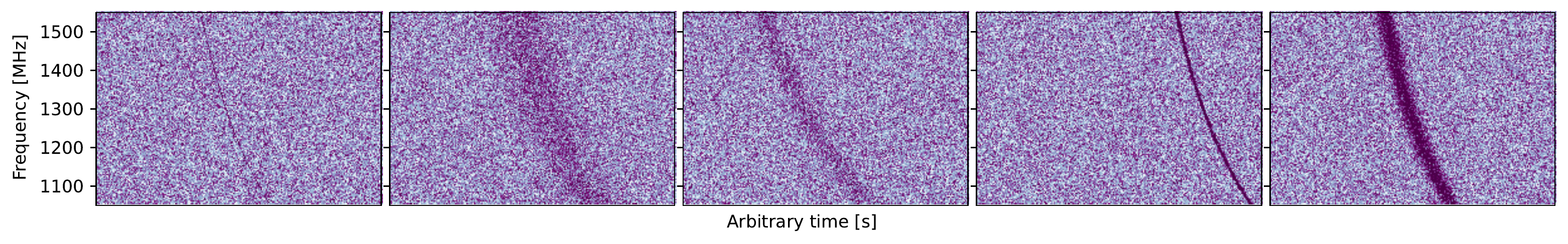}
  \caption{Simple pulses}
  \label{fig:simplepulse_paf_01}
\end{subfigure}
\begin{subfigure}[t]{\textwidth}
  \includegraphics[width=\textwidth]{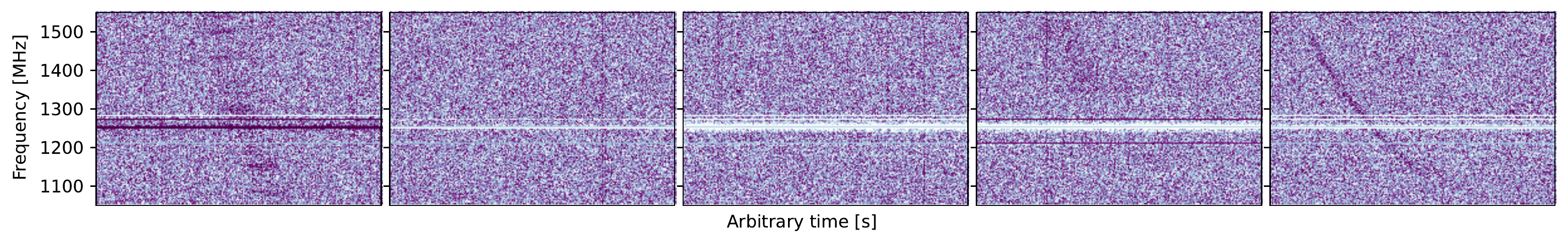}
  \caption{Real pulses}
  \label{fig:realpulse+rfi_paf_01}
\end{subfigure}
\begin{subfigure}[t]{\textwidth}
  \includegraphics[width=\textwidth]{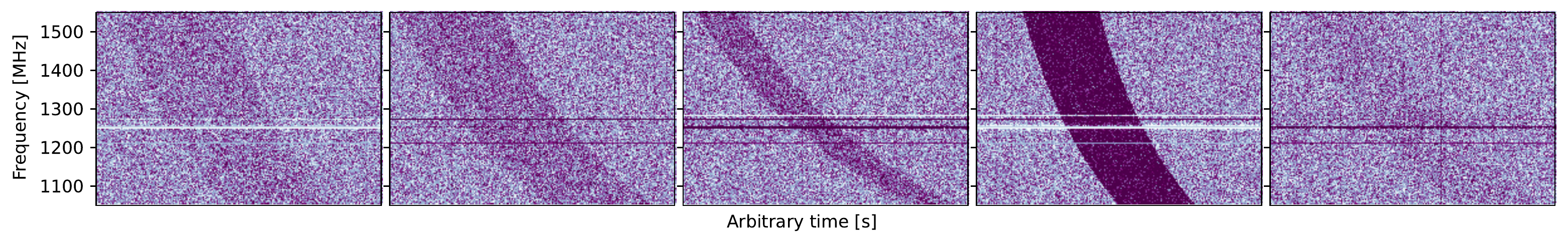}
  \caption{Long pulses}
  \label{fig:longpulse+rfi_paf_01}
\end{subfigure}
\begin{subfigure}[t]{\textwidth}
  \includegraphics[width=\textwidth]{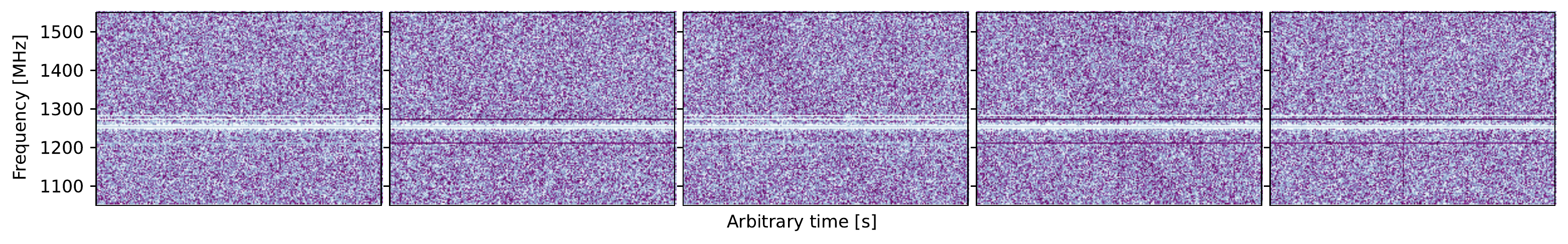}
  \caption{Stellar flares}
  \label{fig:flare+rfi_paf_01}
\end{subfigure}
\begin{subfigure}[t]{\textwidth}
  \includegraphics[width=\textwidth]{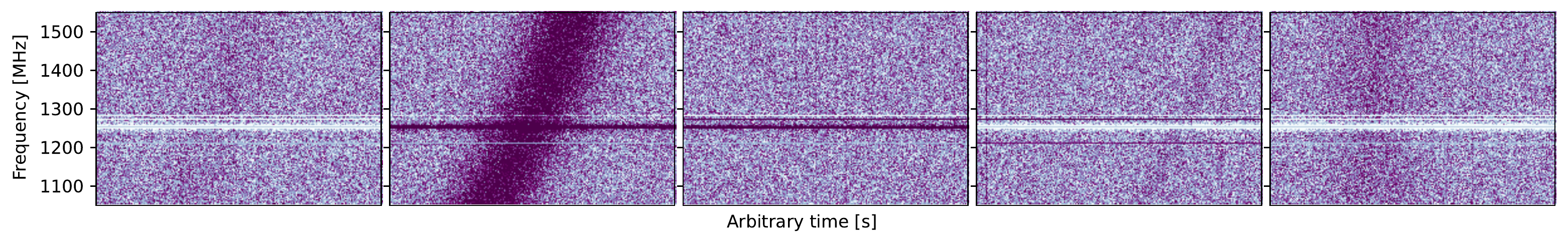}
  \caption{Negative DM pulses}
  \label{fig:ndmpulse+rfi_paf_01}
\end{subfigure}
\begin{subfigure}[t]{\textwidth}
  \includegraphics[width=\textwidth]{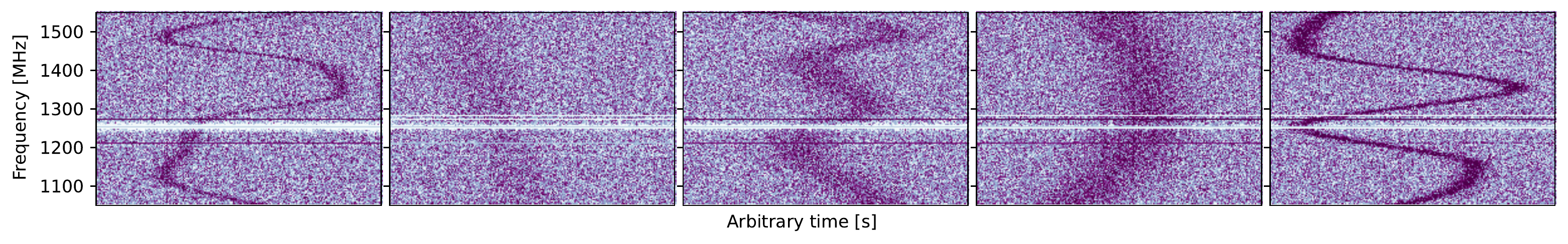}
  \caption{Splines}
  \label{fig:spline+rfi_paf_01}
\end{subfigure}
\begin{subfigure}[t]{\textwidth}
  \includegraphics[width=\textwidth]{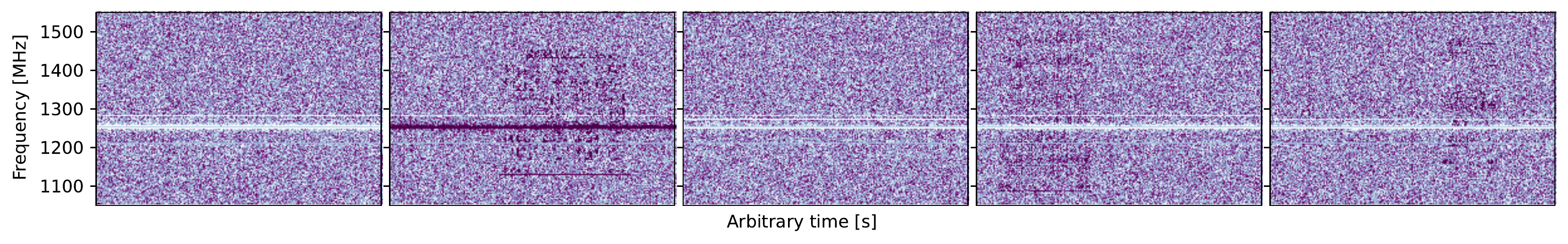}
  \caption{Steganography images}
  \label{fig:image+rfi_paf_01}
\end{subfigure}
\caption{Examples of simulated events for the PAF survey.}
\label{fig:simsignal_paf_01}
\end{figure*}

Distributions of the properties for the simulated and injected signals are provided in \cref{tab:simsignalproperty} and justified in the sections below. Examples of simulated events are displayed in \cref{fig:simsignal_multi_01} for the multibeam and \cref{fig:simsignal_paf_01} for the PAF survey. The examples have been chosen so the events are visually identifiable. In general, due to the range of parameters that we have chosen, most of the simulated signals or events are either too faint and/or too narrow to be seen visually. Note that in the actual observations from the multibeam receiver system, the channel bandwidth is negative and so the highest observing frequency is at the bottom of such plots for the multibeam survey dataset. This is accounted for in our simulated multibeam datasets, but for clarity has been switched-around in the figures presented here.

\begin{table*}
  \centering
  \caption{List of the source types that are either added to simulated radiometer noise or injected into actual observations.}
  \label{tab:simsignalgroup}
  \begin{tabular}{@{\extracolsep{4pt}}l l c@{}}
  \toprule
  Event Group & Type of signals and events & Simulated/Injected \\
  \midrule
  \texttt{noise} & Radiometer noise & Simulated \\
  \texttt{rfi} & Radiometer noise + RFI & Simulated \\
  \texttt{simplepulse} & Simple, idealised, pulses & Simulated \\
  \texttt{known+rfi} & Simple pulses + real pulses + long pulses + flares + RFI & Simulated \\
  \texttt{unknown+rfi} & Negative DM pulses + splines + images + RFI & Simulated \\
  \texttt{combo+rfi} & Known unknowns + unknown unknowns + RFI & Simulated \\
  \texttt{real+combo}$^{a}$ & Real data + known unknowns + unknown unknowns & Injected \\
  \bottomrule
  \multicolumn{3}{p{0.8\linewidth}}{\footnotesize$^{a}$ Only for multibeam survey.} \\
  \end{tabular}
\end{table*}

\begin{table*}
  \centering
  \caption{Property distribution of simulated and injected signals.}
  \label{tab:simsignalproperty}
  \begin{tabular}{@{\extracolsep{4pt}}l l c@{}}
  \toprule
  Type of Signals & Property & Parameter Distribution \\
  \midrule
  \multicolumn{3}{c}{Known unknowns} \\
  \midrule
  Simple and realistic pulses & Amplitude [Jy] & LogUniform($10^{-2}$, 20) \\
   & Width [s] & LogUniform($10^{-4}$, 0.1) \\
   & Dispersion measure [\pccm{}] & Uniform(0, $\sim 1000$)$^{a}$ \\
   & Spectral exponent & Constant(-2) \\
  Long period pulses & Amplitude [Jy] & LogUniform($10^{-2}$, 20) \\
   & Width [s] & LogUniform(0.1, 0.5) \\
   & Dispersion measure [\pccm{}] & Uniform(0, $\sim 1000$)$^{a}$ \\
   & Spectral exponent & Constant(-2) \\
  Stellar flares & Peak flux [Jy] & LogUniform($10^{-2}$, 1) \\
   & Width [s] & LogUniform($10^{-3}$, 0.5) \\
   & Total bandwidth [MHz] & LogUniform(100, 2000) \\
   & Drift rate [MHz\,s$^{-1}$] & $\pm$LogUniform(0.5, 2000) \\
  \midrule
  \multicolumn{3}{c}{Unknown unknowns} \\
  \midrule
  Negative DM pulses & Amplitude [Jy] & LogUniform($10^{-2}$, 20) \\
   & Width [s] & LogUniform($10^{-4}$, 0.1) \\
   & Dispersion measure [\pccm{}] & Uniform($\sim -1000$, 0)$^{a}$ \\
   & Spectral exponent & Uniform(-1.5, 4) \\
  Spline curves & Amplitude [Jy] & LogUniform($10^{-2}$, 20) \\
   & Width [s] & LogUniform($10^{-4}$, 0.1) \\
   & Number of nodes & Uniform(2, 10) \\
  Steganography images & Amplitude [Jy] & LogUniform($10^{-2}$, 100) \\
  \bottomrule
  \multicolumn{3}{p{0.8\linewidth}}{\footnotesize$^{a}$ Limit on dispersion measure, DM, varies for different surveys, 1087\,\pccm for single-beam of multibeam, 3000\,\pccm for 13-beam of multibeam, and 458\,\pccm for PAF.} \\
  \end{tabular}
\end{table*}

\subsubsection{RFI}

RFI is ubiquitous. Observatories such as Parkes are not situated in large-scale radio quiet zones and visitors to the site bring mobile handsets, WiFi devices, etc. Even observatories within radio quiet zones are affected by satellite and aircraft communications. The observations are also affected by broadband, transient signals from impulsive broadband radio sources such as lightning, or from strong impulsive RFI elsewhere in the band saturating the receiver or backend system.

We have chosen representative files from the actual multibeam survey data sets and inspected them, by eye, to infer reasonable properties of the RFI. These observations were primarily affected by point-to-point microwave links above 1433\,MHz, which modulated on time scales of fractions of a second. The number of point-to-point microwave links of 2\,MHz wide was randomly chosen between 1 and 5 at random central frequencies above 1433\,MHz. The microwave links switch on and off on time scales of seconds. Each microwave link is simulated independently. They have randomly chosen ``on'' and ``off'' durations between 0.1 and 3\,seconds. The peak flux amplitude is randomly distributed $0.1 \leq \subtext{S}{peak} [{\rm Jy}] \leq 1$.

Since there are no actual observations yet from the PAF, we examined data from the SUrvey for Pulsars and Extragalactic Radio Bursts \citep[SUPERB;][]{Keane+:2018}, which are 2-bit sampled, as well as observations from the Parkes ultra-wide-bandwidth receiver \citep{Hobbs+:2020}, which covers the frequency range of the PAF between 1050 and 1550\,MHz. This frequency range requires models of satellite interference, in particular the Global Navigation Satellite System (GNSS). The GNSS constellations include the Global Positioning System (GPS) from United States, Global Navigation Satellite System (GLONASS) from Russia, BeiDou Navigation Satellite System from China, and Galileo from Europe. Up to 5 of these are randomly selected with varying peak flux amplitude of $10 \leq \subtext{S}{peak} [{\rm Jy}] \leq 500$ and can come and go on a scale of 1--10\,minutes.

Impulsive RFI is included in our simulations with logarithmic uniform occurrence rate ranging from 0.1 to 1 percent of the total number of samples. The amplitude of this RFI is modelled following a power law distribution and the width is fixed to be 256\,$\mu$s.

\subsubsection{Simple Pulses}

We assume that idealised, single pulses from pulsars or FRB events can be modelled as a Gaussian profile defined by an amplitude, $A$, and a width, $w$. The lower threshold of the amplitude is selected to be about an order of magnitude lower than that can be detected by \textsc{PRESTO}. We simulate a logarithmic uniform distribution in $A$ and $w$ as $10^{-3} \leq A [{\rm Jy}] \leq 20$ and $10^{-4} \leq w [{\rm second}] \leq 0.1$. Note that these are not designed to represent an actual, physical distribution of these parameters; instead they are chosen to cover the expected parameter space that will be probed during such simulations.

\subsubsection{Known Unknowns}

We consider possible signals that have durations less than a simulated frame ($\sim 1$\,second), are detectable in the frequency band covered, but are not commonly searched for in such data sets.

\paragraph{More realistic single pulses:}
We allow the single pulses from pulsars and FRBs to contain frequency structure. Pulsars exhibit scintillation, whereas FRBs exhibit complex, narrowband frequency structure that is currently unexplained \citep[e.g.,][]{Pleunis+:2021,Dai+:2022}. We model such signals using dynamic spectra with varying intensity at each frequency channel. We randomly choose between a dynamic spectrum based on Kolmogorov turbulence in the interstellar medium and a dynamic spectrum with sharp frequency cutoffs \citep{Coles+:2010,Dai+:2016}. In the latter case, we assume the pulse is only present in only segment of the dynamic spectrum, which is at least 10 frequency channels wide.

\paragraph{Long-duration pulses:}
Pulsar and magnetar-like objects are now known with pulse periods of 10\,s of seconds to several minutes \citep[e.g.,][]{Tan+:2018,Hurley-Walker+:2022}. We therefore simulate single pulse events that follow the dispersion law, but with pulse widths significantly larger than those typically found in pulsar searches. Here we model such events with pulse widths from 0.1 to 0.5\,seconds, modelled as a rectangular profile instead of Gaussian to prevent wide pulses overlapping multiple frames.

\paragraph{Stellar radio bursts:}
Traditional search pipelines explicitly search for dispersed signals, with the pulse frequency drift having a frequency dependence of $\nu^{-2}$. However, stellar radio bursts (such as from brown dwarfs, M-dwarfs, and magnetic chemically peculiar stars) typically exhibit linear frequency drifts in the frequency ranges and time-scales of interest in this work \citep[e.g.,][]{Osten+Bastian:2008,Zic+:2019}, although more complex morphologies are possible \citep[e.g.,][]{Hess+Zarka:2011,Leto+:2017}. Furthermore, the frequency drift rate $\dot \nu$ can also be positive or negative, i.e., the burst may arrive later at higher ($\dot \nu > 0$) or at lower observing frequencies ($\dot \nu < 0$), with respect to a chosen reference frequency \subtext{\nu}{ref}. We model such events with peak flux intensity of $10^{-2} \leq \subtext{S}{peak} [{\rm Jy}] \leq 1$ and Gaussian width $10^{-3} \leq w [{\rm second}] \leq 0.5$.

Although the range of drift rate is set to be $0.5 \leq |\dot{\nu}| [{\rm MHz\,s}^{-1}] \leq 2000$, the value is determined based on the randomly generated event time and \subtext{\nu}{ref} that fits inside a single time frame. The total bandwidth spans $100 \leq \Delta \nu [{\rm MHz}] \leq 2000$. We further apply the constraint that fast drift rate $\dot{\nu} > |\pm 500|\,$MHz\,s$^{-1}$ flares typically have narrow width $w < 0.1\,$s, while those that are slow have wider width. Consequently, mostly, if not all, are fast and narrow flares simulated than slow and broad flares, as the latter has steeper linear drift rate and less likely to fit in the frame. Additionally, fast narrow flares can have multiple components and we simulate 1 to 10 multiple bursts in the same frame with the same drift rate and width, while other parameters vary.

\subsubsection{Unknown Unknowns}

Simulating ``unknown unknowns'' is clearly impossible in a well-defined, quantitative manner. Instead we have searched the literature for likely signatures and attempted to make models of generic functional forms.

\paragraph{Negative DM pulses:}
Pulsars and FRBs are modelled with a positive DM value and spectral exponent of $-2$. Here, we generalise to negative DMs using the same DM limit as minimum, i.e., from $-1087$\,\pccm for multibeam and $-458$\,\pccm for PAF to 0\,\pccm. The exponents range from $-1.5 < \alpha < 4$ with a constraint that the event has to fit within the frame.

\paragraph{Spline curves:}
Splines provide the means to simulate smooth, wide-band, impulsive events of arbitrary form. We randomly inject spline curves with amplitudes from $10^{-2}$ to 20\,Jy, number of nodes from 2 to 10, and a maximum possible width range of $10^{-4}$ to 0.1\,second. The connected nodes are evenly separated in frequency space to yield a smooth curve.

\paragraph{Steganography:}
We can embed information (a message) into the data set. To do this, we can simply convert an image into time-frequency pixels, in a similar fashion to the interstellar radio messages that have been transmitted from Earth. Among those transmitted pictorial messages, we use Arecibo message \citep{Staff+Ionosphere:1975} and Cosmic Call as input images. The Arecibo message contains seven components, which represents numbers, DNA elements, nucleotides, double helix, humanity, planets, and telescope. The Cosmic Call consists of 23 images, with symbols designed to encompass concepts on numbers, mathematics, units, chemistry, physics, biology, and astronomy.

We include 31 steganography images, eight from the Arecibo message, where seven of which are the different components and one is the entire image, and 23 from Cosmic Call. The images are randomly selected and scaled to various sizes with intensity range of $10^{-2}$ to 100\,Jy. The height in pixel is set to be at least half the total number of channels, while the width can be as narrow as 1 pixel.



\subsection{Injection into Real Data} \label{ssec:injectintoreal}

\begin{figure*}
\centering
\includegraphics[width=\textwidth]{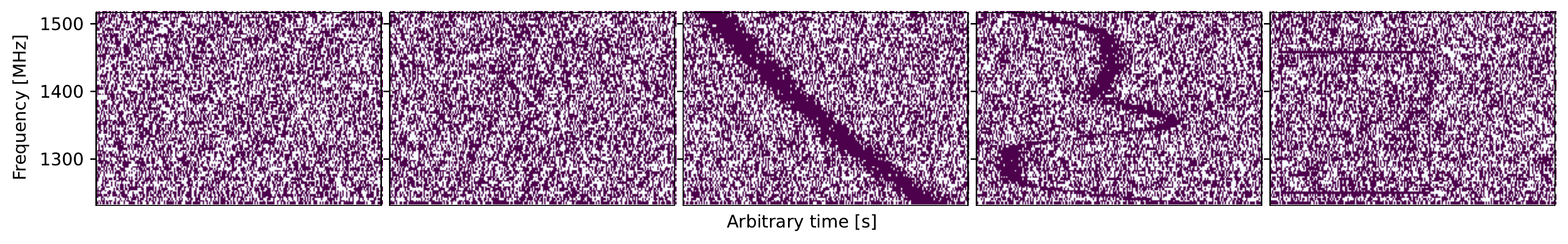}
\caption{Examples of randomly injected events into \texttt{PM0001\_00111.sf} multibeam file.}
\label{fig:real+combo_multi_01_PM0001_00111}
\end{figure*}

All combinations of the simulated signals are also injected into actual observational data for multibeam surveys. No extra radiometer noise nor RFI has been added into these. We randomly chose 50 real observation data files from the multibeam survey, which were downloaded from the CSIRO data archive. Each file has 2051 sub-integrations. We injected into 10 percent of the total frames an event (corresponding to 205 events). Examples of injected signals into real data are shown in \cref{fig:real+combo_multi_01_PM0001_00111}. Of course the real data contains RFI and potentially contains astronomical sources. We explicitly chose files which do not contain known pulsars or FRBs events.

For the 13-beam data sets, we specifically chose the set of 13 files that contain a well-known FRB, the Lorimer burst, which was downloaded from the CSIRO DAP\footnote{\url{https://doi.org/10.4225/08/5819628e4fed9}}. The actual Lorimer burst is detected in sub-integration 421 primarily in beams 6 and 13. The sampling time is higher compared to the earlier surveys and so the maximum DM limit is set for our injected events to be 3000\,\pccm. Similarly, 10 percent of the total 2051 sub-integrations are injected with 205 events. This dataset is not partitioned into frames as this allows us to test streaming-based search methods. The signal strength in a particular beam depends upon the beam pattern (which we model using a simple sinc$^{2}$-function) and the distance from the source to the beam centre. Clearly our simulations would not be useful if the events were all simulated well-away from the beam pointing direction and so we provide three different scenarios for the source position, which are illustrated in \cref{fig:real+combo_multimb_SMC021_008} and detailed below:
\begin{enumerate}
\renewcommand{\theenumi}{(\arabic{enumi})}
  \item Type 1: For a given event, we randomly choose which beam it will be in and then inject it as if the source was in the pointing direction of that beam. The event is not injected into the other beams. In \cref{fig:real+combo_multimbt1_SMC021_008}, the source is injected into the central beam.
  \item Type 2: Each event is always assumed to come from the pointing direction of the central beam, but the signal is also seen in the other beams with the amplitude being scaled by the angular offset of that beam to the source. So, in this case, the central beam contains all the events exactly as they are simulated and the other beams also contains the same event, but at weaker amplitudes. In \cref{fig:real+combo_multimbt2_SMC021_008}, the source is injected into the central beams, but the signal strength is so strong that it is detected also in the side-lobes of the outer beams.
  \item Type 3: For every event, we randomly choose a sky position in the field of view and so sometimes the event will be strongest in one beam, sometimes in another and sometimes half-way between two. In \cref{fig:real+combo_multimbt3_SMC021_008}, the source position is slightly offset from the central beam position.
\end{enumerate}

\begin{figure*}
\centering
\begin{subfigure}[t]{\textwidth}
  \includegraphics[width=\textwidth]{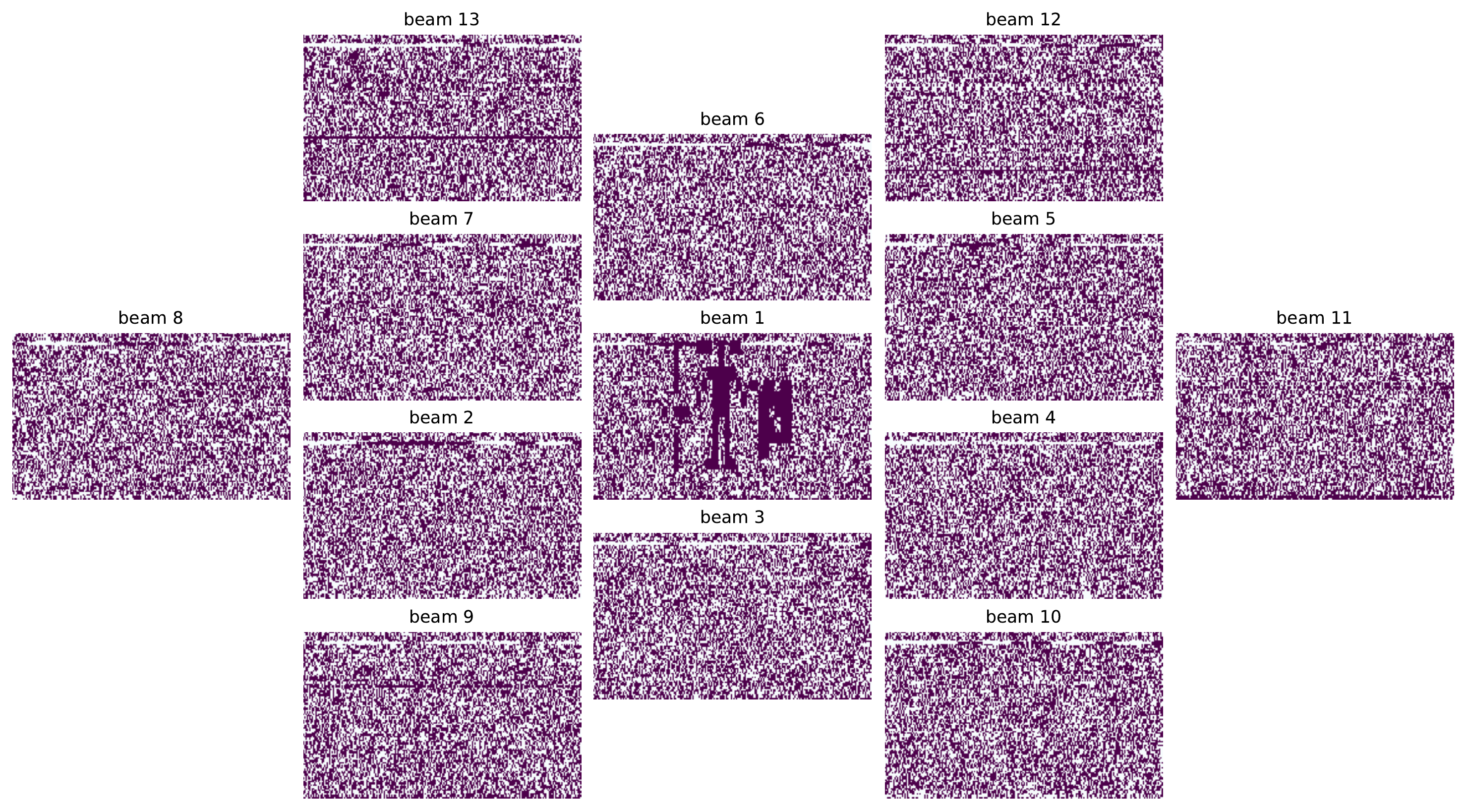}
  \caption{Type 1: Random injection into one beam}
  \label{fig:real+combo_multimbt1_SMC021_008}
\end{subfigure}
\begin{subfigure}[t]{\textwidth}
  \includegraphics[width=\textwidth]{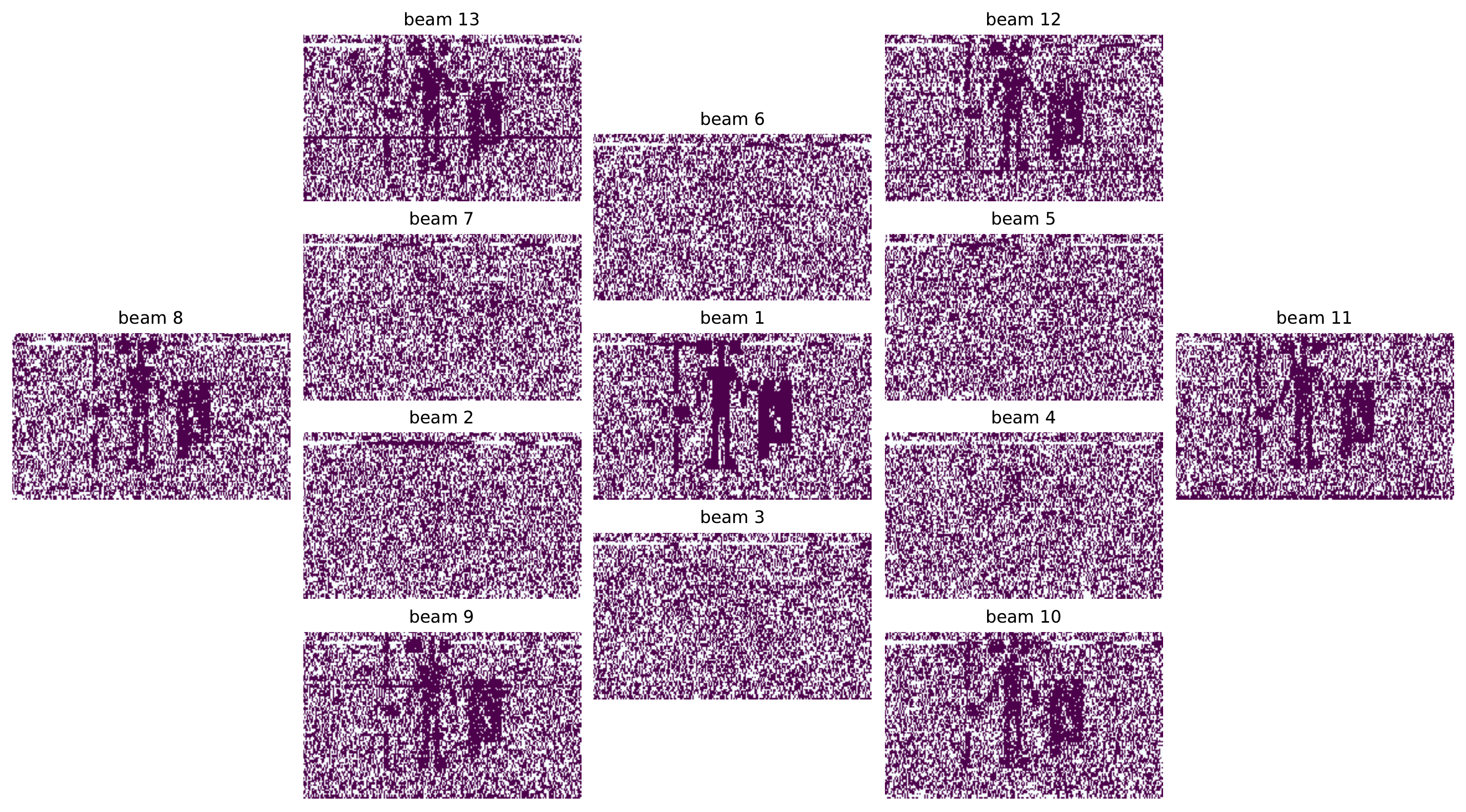}
  \caption{Type 2: Strongest amplitude in the central beam while weaker for the rest of the beams (Note that as the signals are strong, they are in the outer beams due to the side-lobe modelling)}
  \label{fig:real+combo_multimbt2_SMC021_008}
\end{subfigure}
\caption{Examples of an injected event into all beams of the \texttt{SMC021\_008*.sf} file. \label{fig:real+combo_multimb_SMC021_008}}
\end{figure*}%
\begin{figure*}
\ContinuedFloat
\begin{subfigure}[t]{\textwidth}
  \includegraphics[width=\textwidth]{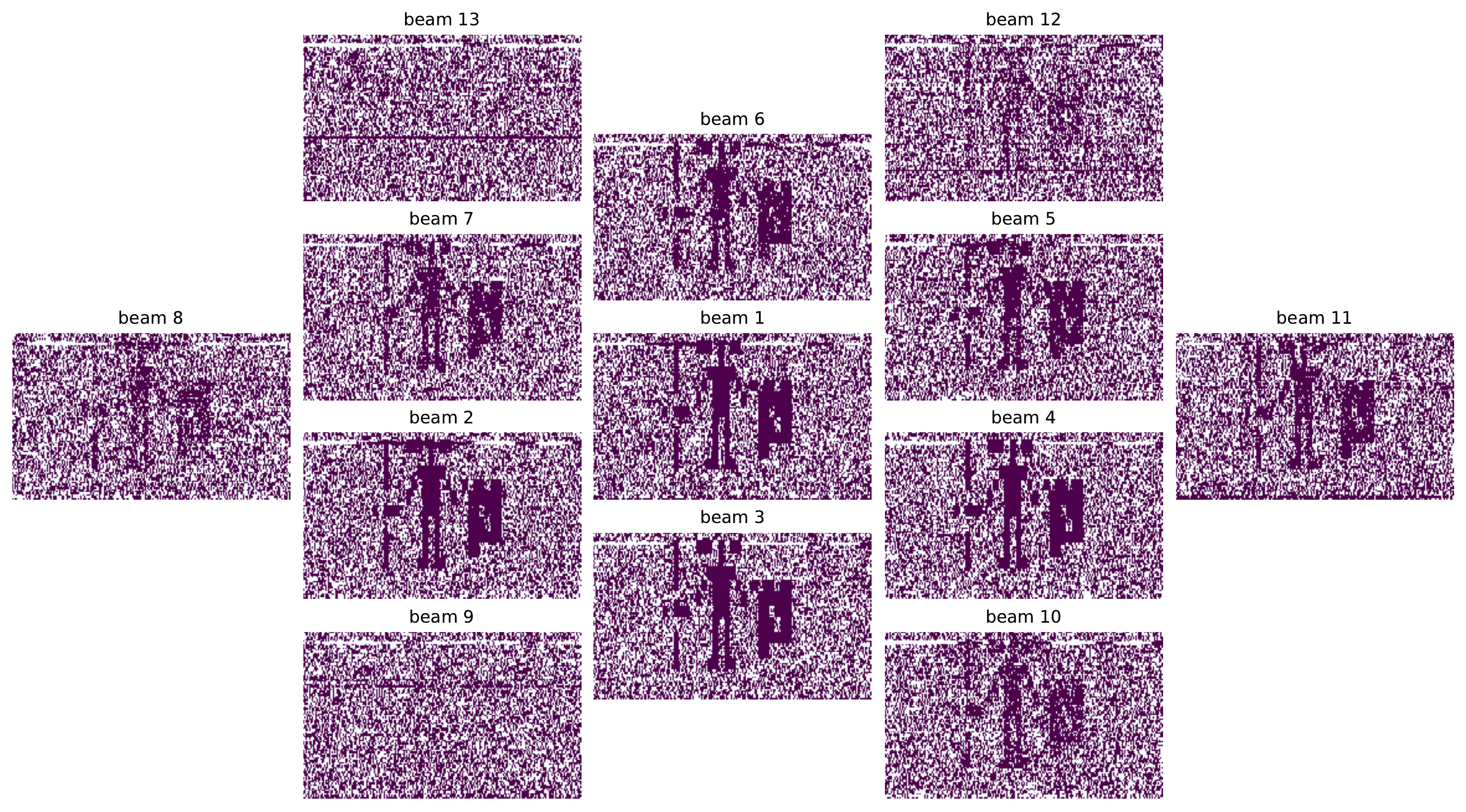}
  \caption{Type 3: Randomly chosen source position within 1\,degree of the central position}
  \label{fig:real+combo_multimbt3_SMC021_008}
\end{subfigure}
\caption{--- Continued \label{fig:real+combo_multimb_SMC021_008}}
\end{figure*}

\section{Applying \textsc{PRESTO} to the Data Challenge} \label{sec:searchpresto}

Our data challenge consists of a large number of labelled events. The primary use of this challenge will be to compare the effectiveness of different algorithms in detecting the simulated signals. A commonly-used software package for searching for signals affected by cold-plasma dispersion such as bright periodic pulses and FRBs, is \textsc{PRESTO} \citep{Ransom:2001,Ransom+:2002}. \textsc{PRESTO} is used to de-dispserse the data files at a range of DMs, produce time series for each DM trial by averaging across the frequency channels (accounting for the presence of RFI and bad channels) and then searching for pulsed signals. In the following analysis, we restrict our analysis to the single beam mock surveys as the primary single-pulse search algorithm in \textsc{PRESTO} runs on single beam data sets. We note that the output candidates could (and often are) subsequently shifted based on multibeam information.

\textsc{PRESTO} has many options to optimise the search for a particular survey. Here, we choose a simple set of parameters (mostly using the default values) as representative \textsc{PRESTO} output. We therefore expect that our \textsc{PRESTO}-based pipeline to be optimal for simple FRB-like pulses in a low RFI environment. Our pipeline is not expected to be optimal for other types of events.

The outline of this search pipeline is as follows:
\begin{enumerate}
\renewcommand{\theenumi}{(\arabic{enumi})}
  \item Process each search-mode data file independently.
  \item Mask RFI using \texttt{rfifind}, except for data files only containing radiometer noise. The integration time is set to be 2\,second. The ``no scales'' and ``no offsets'' options are also used.
  \item Create a de-dispersion plan using \texttt{DDplan.py}. The lower limit of DM search range is set to be 0\,\pccm and upper DM of 800\,\pccm for the multibeam and 300\,\pccm for the PAF survey. This upper limit is purposely set to be lower than the simulated DM values to probe the drop off in detection rate for events at high DM. The observation parameters for \texttt{DDplan.py} include the central frequency, bandwidth, number of channels, and sampling time, which are based on the raw \texttt{PSRFITS} file header information.
  \item Run de-dispersion on the data using \texttt{prepsubband} from the generated \texttt{DDplan.py}. The number of subbands is set to 32. The ``no scales'', ``no offsets'', ``no barycenter'', and ``no clip'' options are also used.
  \item Search for single pulses in the data with a boxcar matched filter using \texttt{single\_pulse\_search.py}. The ``fast'' and ``no bad-blocks'' options are enabled (noting that this will reduce \textsc{PRESTO}'s sensitivity to long-duration events). A signal-to-noise (S/N) threshold of 7$\sigma$ is used to identify candidates that get recorded to disk for comparison with the injected events.
\end{enumerate}

The single pulse search yields output files containing a list of candidates along with their estimated DM, S/N, and time. We remove candidates with zero DM. The resulting candidates are then binned per frame and those with the highest S/N within that bin are then returned. This ensures that multiple detections of the same event are grouped together and only counted once.

We note that there are differences between the processing of the two simulated surveys. Our \textsc{PRESTO} single pulse search sets the maximum down-sampling to be 32 times of the sampling time, which is 8 and 1.8\,ms respectively for the multibeam and PAF surveys. This choice of default parameterisation therefore means that we are less sensitive to wide events in the PAF survey than in the multibeam survey.

\subsection{Baseline Scoring Metrics}

The goal is to maximise the chance of finding astronomical signals in the datasets, and so the relevant evaluation metrics reflect the number of true and false positives and negatives found. These are presented in the confusion matrix shown in \cref{tab:cmevalmetric}. The true positive (TP) and true negative (TN) values are the numbers of the correctly predicted outcomes for events and non-events, respectively. False positive (FP) and false negative (FN) events are those that are wrongly predicted as an event and non-event, respectively. The total positives (P) is the sum of TP and FN, while the total negatives (N) is the sum of TN and FP. 

\begin{table*}
  \centering
  \caption{Confusion matrix and evaluation metrics for different event groups using \textsc{PRESTO}. The evaluation metrics scores range from 0 to 1, with 1 being perfect, except for fall-out where 0 is best.}
  \label{tab:cmevalmetric}
  \begin{tabular}{l @{\hspace{0.8cm}} *{7}{c} @{\hspace{0.8cm}} *{4}{c}}
  \toprule
  Event Group & \multicolumn{7}{c}{Confusion Matrix} & \multicolumn{4}{c}{Evaluation Metrics} \\
  \cmidrule(lr){2-8} \cmidrule(lr){9-12} \\
   & N & P & N+P & FN & FP & TN & TP & Recall & Precision & Fall-out & F1 Score \\
  \midrule
  \multicolumn{12}{c}{multibeam} \\
  \midrule
  noise & 50000 & 0 & 50000 & 0 & 0 & 50000 & 0 & 0 & 0 & 0 & 0 \\
  rfi & 50000 & 0 & 50000 & 0 & 0 & 50000 & 0 & 0 & 0 & 0 & 0 \\
  simplepulse & 45000 & 5000 & 50000 & 2346 & 320 & 44680 & 2654 & 0.5308 & 0.8924 & 0.0071 & 0.6657 \\
  known+rfi & 45000 & 5000 & 50000 & 3216 & 1572 & 43428 & 1784 & 0.3568 & 0.5316 & 0.0349 & 0.4270 \\
  unknown+rfi & 45000 & 5000 & 50000 & 3728 & 1510 & 43490 & 1272 & 0.2544 & 0.4572 & 0.0336 & 0.3269 \\
  combo+rfi & 45000 & 5000 & 50000 & 3408 & 1680 & 43320 & 1592 & 0.3184 & 0.4866 & 0.0373 & 0.3849 \\
  real+combo & 92300 & 10250 & 102550 & 7212 & 2709 & 89591 & 3038 & 0.2964 & 0.5286 & 0.0293 & 0.3798 \\
  \midrule
  \multicolumn{12}{c}{PAF} \\
  \midrule
  noise & 50000 & 0 & 50000 & 0 & 6 & 49994 & 0 & 0 & 0 & 0.0001 & 0 \\
  rfi & 50000 & 0 & 50000 & 0 & 280 & 49720 & 0 & 0 & 0 & 0.0056 & 0 \\
  simplepulse & 45000 & 5000 & 50000 & 2548 & 410 & 44590 & 2452 & 0.4904 & 0.8567 & 0.0091 & 0.6238 \\
  known+rfi & 45000 & 5000 & 50000 & 3104 & 1228 & 43772 & 1896 & 0.3792 & 0.6069 & 0.0273 & 0.4668 \\
  unknown+rfi & 45000 & 5000 & 50000 & 3221 & 1101 & 43899 & 1779 & 0.3558 & 0.6177 & 0.0245 & 0.4515 \\
  combo+rfi & 45000 & 5000 & 50000 & 3146 & 1084 & 43916 & 1854 & 0.3708 & 0.6310 & 0.0241 & 0.4671 \\
  \bottomrule
  \end{tabular}
\end{table*}

From the confusion matrix, we can evaluate various scoring metrics including the recall, precision, fall-out, and F1 score. 
\begin{itemize}
  \item Recall (also known as sensitivity, hit rate, or true positive rate) measures the ability to identify all positives, which is given by
  \begin{equation*}
  \text{Recall}=\frac{\text{TP}}{\text{TP}+\text{FN}}.
  \end{equation*}
  \item Precision (or positive predictive value) measures the ability to not mislabel negatives as positives, which is given by
  \begin{equation*}
  \text{Precision}=\frac{\text{TP}}{\text{TP}+\text{FP}}.
  \end{equation*}
  \item Fall-out (or false positive rate) measures the probability of mislabelled positives that are negatives, which is given by
  \begin{equation*}
  \text{Fall-out}=\frac{\text{FP}}{\text{FP}+\text{TN}}.
  \end{equation*}
  \item F1 score measures the balance between precision and recall, which is given by
  \begin{equation*}
  \text{F1}=2\left(\frac{1}{\text{Recall}} + \frac{1}{\text{Precision}}\right)^{-1}.
  \end{equation*}
\end{itemize}
The F1 score is used since it is less biased toward data sets where there are more events than non-events \citep{He+Ma:2013}.

These scoring metrics are presented in \cref{tab:cmevalmetric}. All scores range from 0 to 1, with 1 being perfect, except for fall-out, in which a lower score represents a better result. 


\subsection{Results using \textsc{PRESTO}}

The primary goals for running \textsc{PRESTO} on our simulation are (1) to demonstrate that the simulations are reliable and (2) to ensure that we cover sufficient parameter space in the properties of the injected signals to allow us to compare the effectiveness of new algorithms with \textsc{PRESTO}.

The simplest simulation contains only radiometer noise. As shown in \cref{tab:cmevalmetric}, there were no false detections using \textsc{PRESTO} for the multibeam simulation, but six for the PAF simulation. The six false detections had S/N $< 7.4\sigma$, which is only slightly above our cut-off S/N of 7$\sigma$. We did not explicitly choose the cut-off based on the expected false alarm rate (which is commonly determined using simulations such as our data set) and therefore this is not unexpected.

We expected that when RFI is added into the radiometer noise the number of false detections would increase. This does occur for the PAF simulations, but not for the multibeam. Both simulated surveys include impulsive RFI, which has a width of 256\,$\mu$s equalling the sampling rate for the multibeam survey, but contains multiple samples for the PAF sampling. The majority of the PAF-survey false detections were caused by impulsive transient events and detected with very low DM values (DM = 0.05\,\pccm, noting that we explicitly did not search candidates with DM = 0\,\pccm).

\begin{figure*}
\centering
\begin{subfigure}[t]{\textwidth}
  \includegraphics[width=\textwidth]{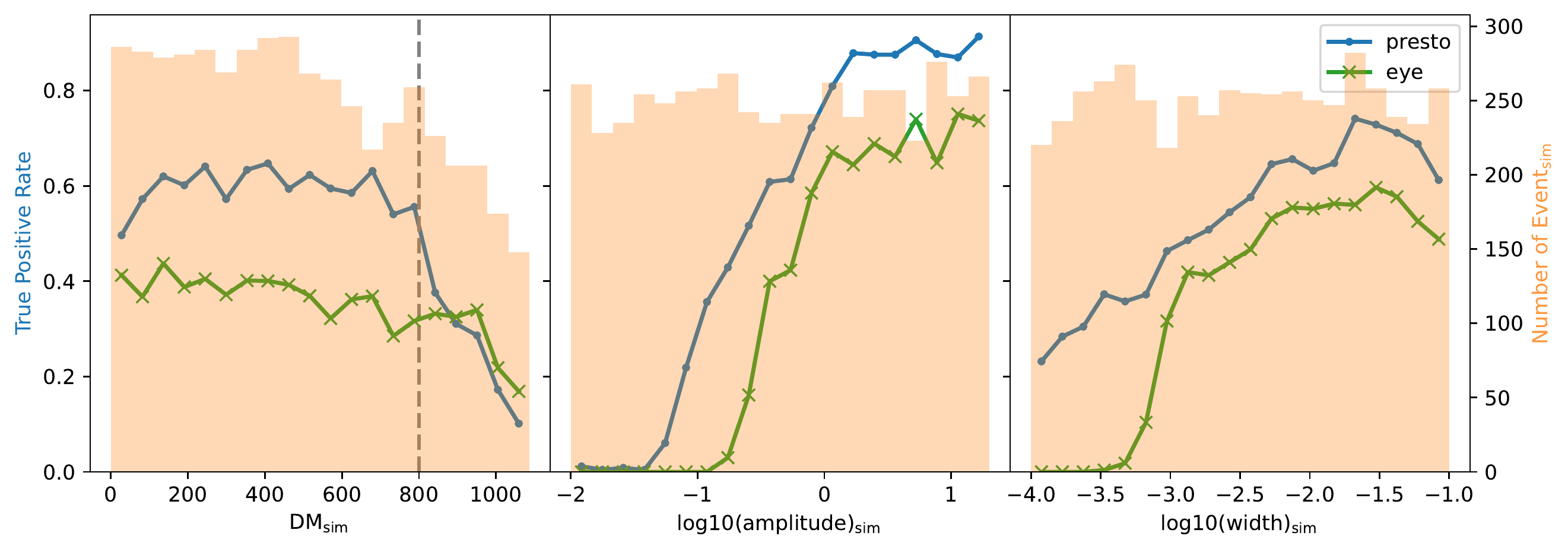}
  \caption{For single-beam using \textsc{PRESTO} and by eye}
  \label{fig:truepositiverate_simplepulse_multi_eye}
\end{subfigure}
\begin{subfigure}[t]{\textwidth}
  \includegraphics[width=\textwidth]{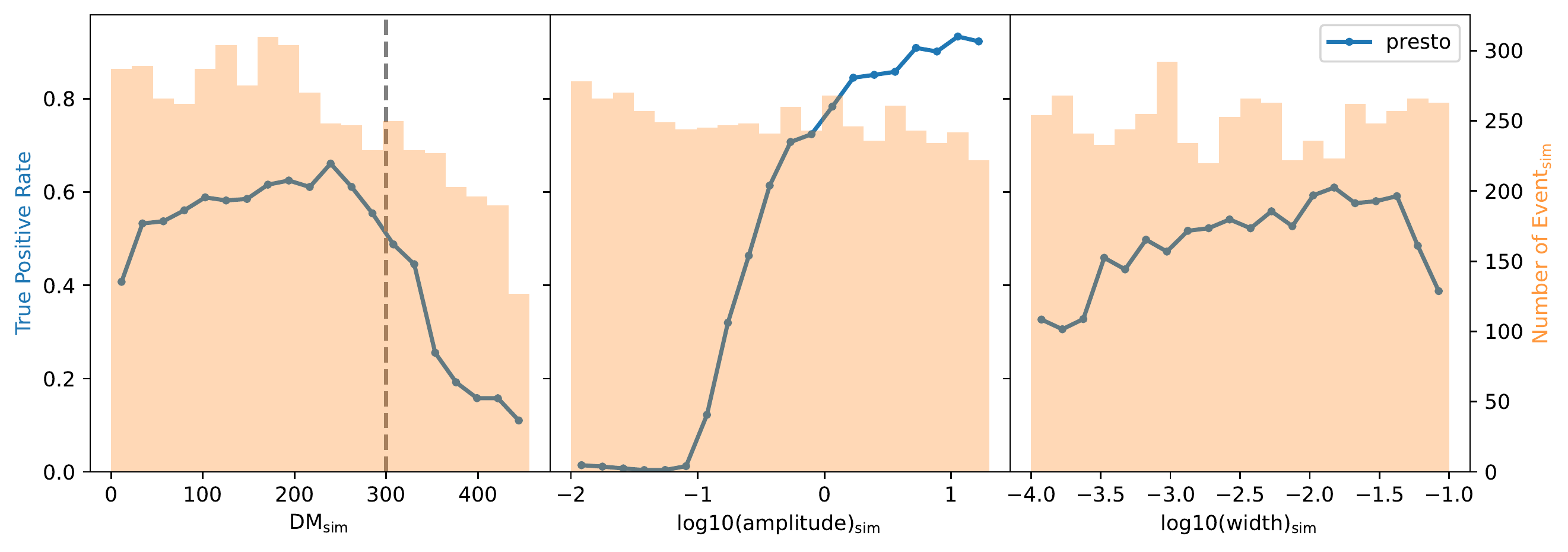}
  \caption{For PAF using \textsc{PRESTO}}
\end{subfigure}
\caption{True positive rate as a function of dispersion measure ({\em left}), amplitude ({\em centre}), and width ({\em right}) for simulated simple pulses. The grey dashed vertical line indicates the DM search range by \textsc{PRESTO}, 0--800\,\pccm for single-beam and 0--300\,\pccm for PAF.}
\label{fig:truepositiverate_simplepulse}
\end{figure*}

\begin{figure*}
\centering
\begin{subfigure}[t]{0.6\textwidth}
  \includegraphics[width=\textwidth]{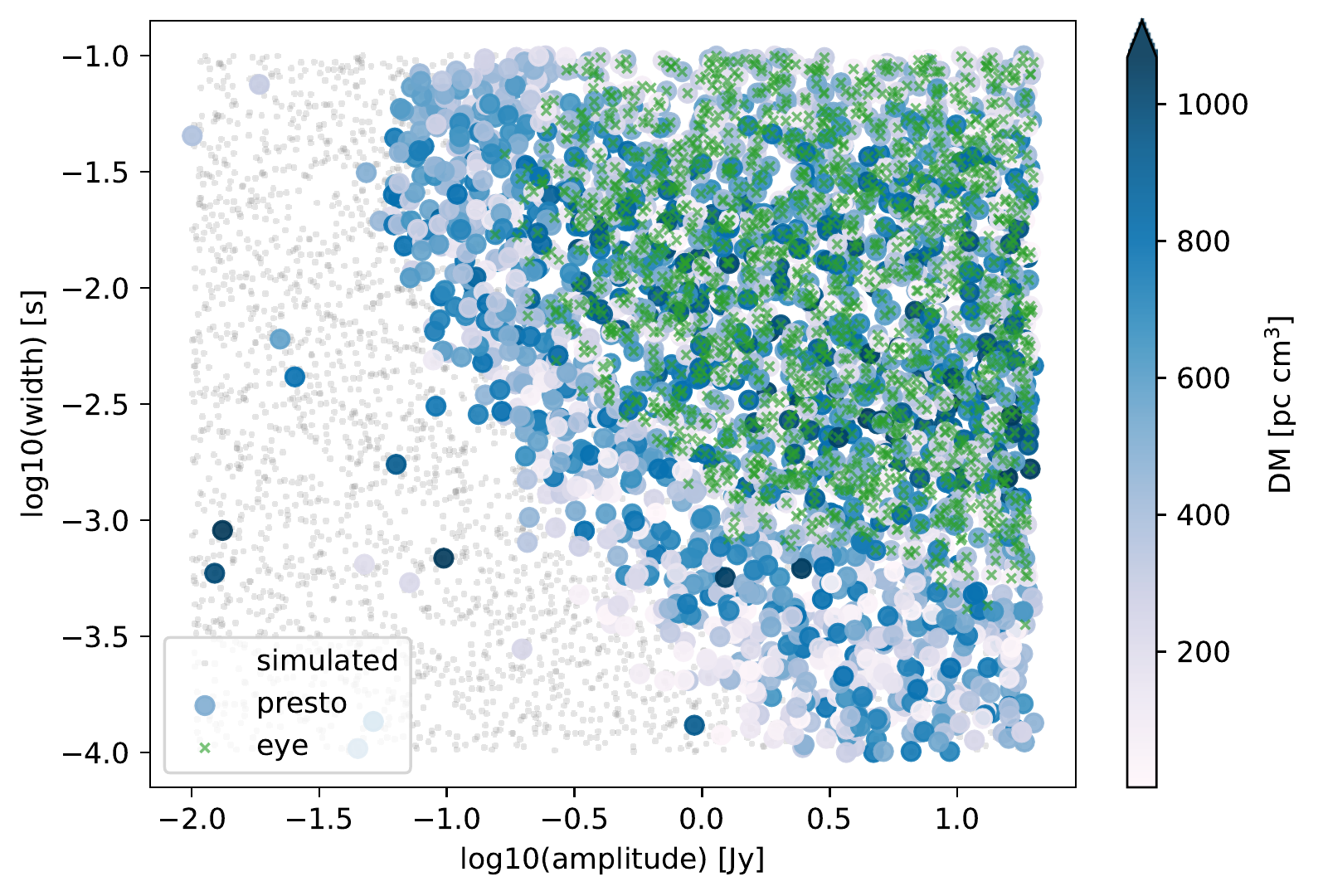}
  \caption{For single-beam using \textsc{PRESTO} and by eye}
  \label{fig:amplitude+width+dm_simplepulse_multi_eye}
\end{subfigure}
\begin{subfigure}[t]{0.6\textwidth}
  \includegraphics[width=\textwidth]{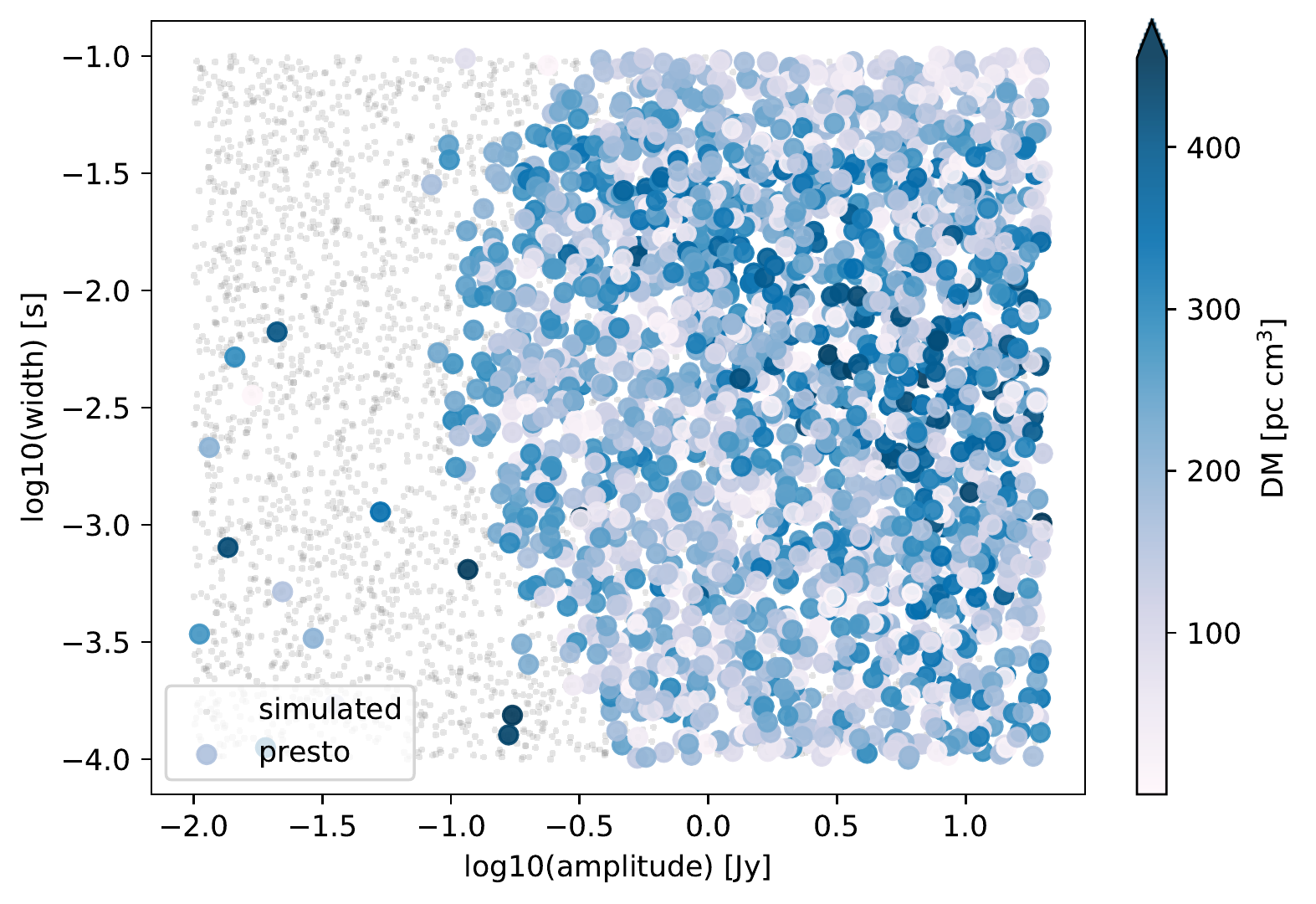}
  \caption{For PAF using \textsc{PRESTO}}
\end{subfigure}
\caption{Detectability of simulated simple pulses in the amplitude, width, and dispersion measure parameter space.}
\label{fig:amplitude+width+dm_simplepulse}
\end{figure*}

The \texttt{simplepulse} dataset, which contains idealised, single pulse events, without any RFI, represents the scenario where \textsc{PRESTO} is expected to be optimal. \Cref{fig:truepositiverate_simplepulse} shows the true positive detection rate (blue solid line). \cref{fig:amplitude+width+dm_simplepulse} plots the detectability of simulated events as a function of amplitude, width and dispersion measure. The orange histogram in each panel of \cref{fig:truepositiverate_simplepulse} shows the injected distribution of event parameters. For the pulse amplitudes and widths, the injected amplitude is flat. However, the number of generated high-DM pulses reduces with increasing DM because of the requirement of ensuring that each event fits within a single frame.

We searched for single pulse events with DMs up to 800 and 300\,\pccm for the multibeam and PAF surveys, respectively. We therefore expect (and see) a drop-off in the detectability of single pulse events higher than those cut-off values, as indicated by the grey dashed vertical line in the left panels of \cref{fig:truepositiverate_simplepulse}. As shown in the central panels (where we show the true positive rate as a function of amplitude), we have chosen a range of amplitudes that covers from being undetectable with \textsc{PRESTO} to close to 100 percent detection. This allows users of this data challenge to compare their own algorithms against \textsc{PRESTO} in low, mid, and high-S/N cases. We note that comparing the central panels for the two simulated surveys gives the impression that the simulated PAF survey is less sensitive than the simulated multibeam survey. As described in \cref{sec:searchpresto}, this is caused by our choice of \textsc{PRESTO} parameters means (we are less sensitive to wide events in the PAF survey compared with the multibeam survey). In terms of the evaluation metrics (\cref{tab:cmevalmetric}), \textsc{PRESTO} provides baseline measurements of the four parameters. In summary, the F1 score of $\sim 0.6$ and precision of $\sim 0.9$ shows that \textsc{PRESTO} is, as expected, an effective algorithm for finding ideal, simple, single pulse events.

Our procedure does produce a small number of false ``true positives''. For instance, in \cref{fig:amplitude+width+dm_simplepulse} we see true positive detections for a few very low amplitude events. This can occur simply by chance (i.e., a $7\sigma$ result in the frame corresponding to a weak event, and hence, identified as a ``true positive''). This can also occur if a bright single pulse was simulated near an edge of a frame. Searching at an incorrect DM can smear some of that bright signal into the adjacent frame where a weaker (and undetectable signal was simulated) and that earlier frame is identified incorrectly as a ``true positive''. For this analysis we ignore this issue and note that a more painstaking analysis of the \textsc{PRESTO} candidates could resolve these issues. 

\begin{figure*}
\centering
\begin{subfigure}[t]{0.33\textwidth}\vskip 0pt
  \imagebox{6cm}{\includegraphics[width=\textwidth,height=6cm]{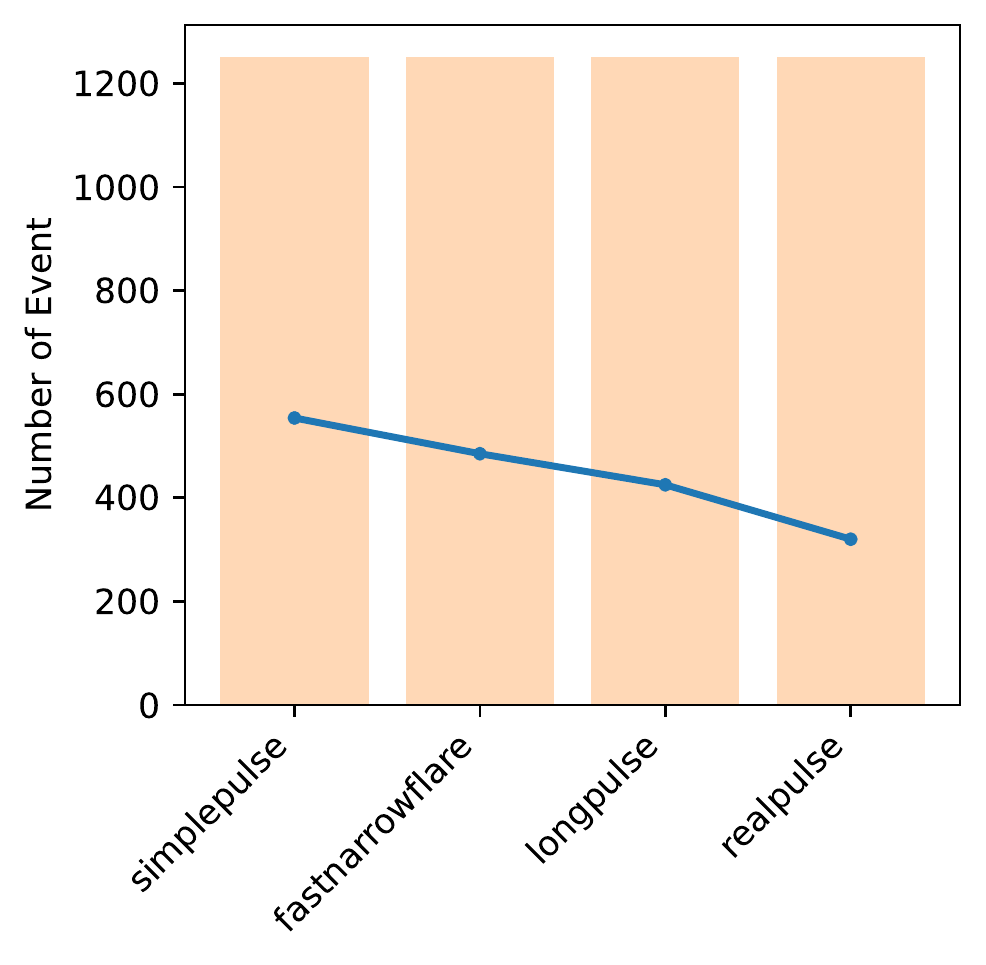}}
  \caption{known+rfi}
\end{subfigure}%
\begin{subfigure}[t]{0.33\textwidth}\vskip 0pt
  \imagebox{6cm}{\includegraphics[width=\textwidth]{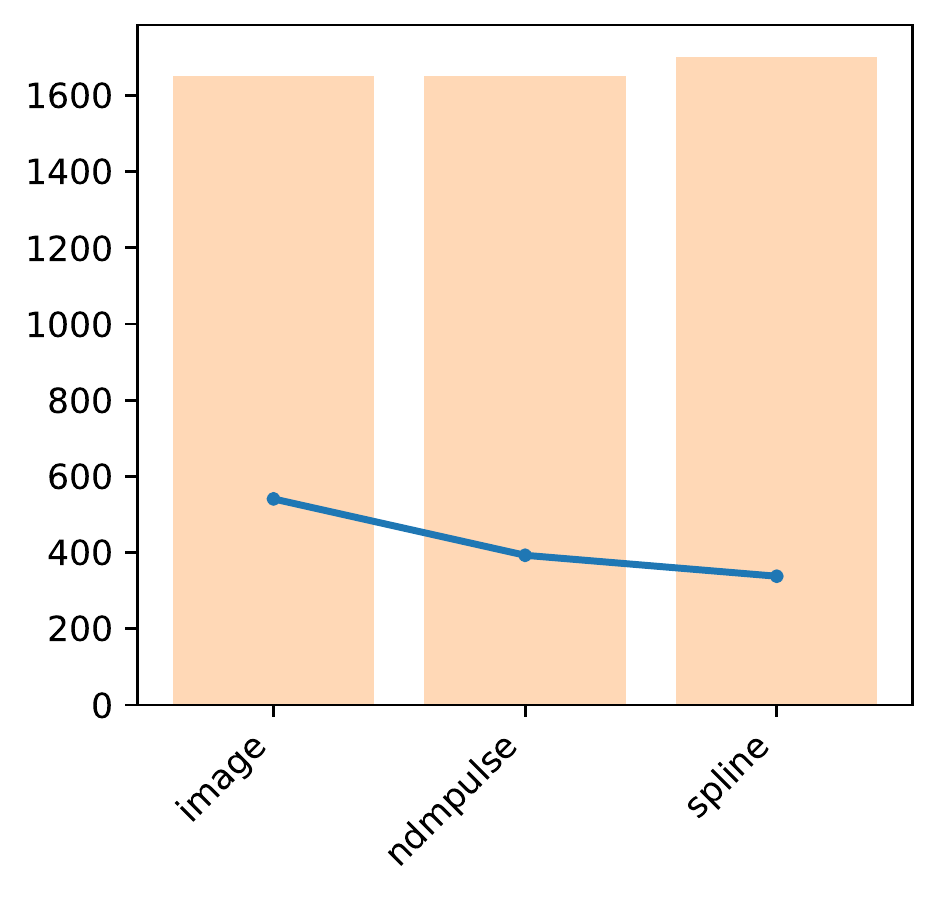}}
  \caption{unknown+rfi}
  \label{fig:tpcount_unknown_multi}
\end{subfigure}%
\begin{subfigure}[t]{0.33\textwidth}\vskip 0pt
  \imagebox{6cm}{\includegraphics[width=\textwidth]{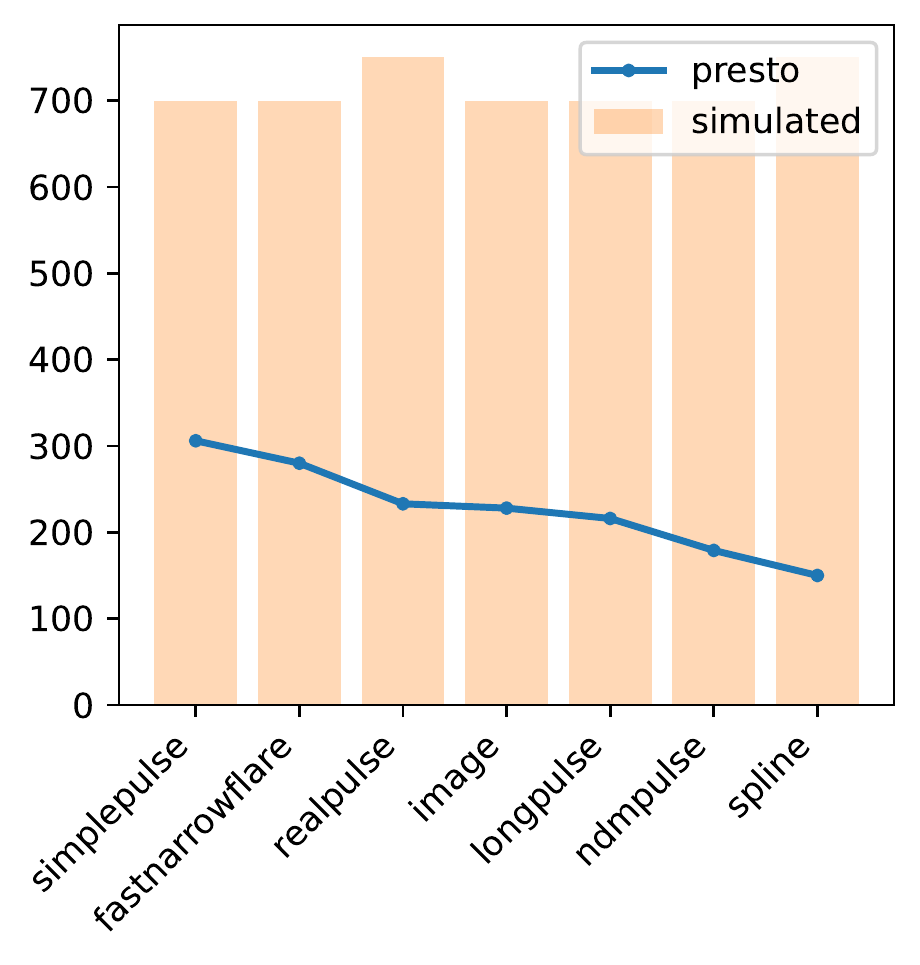}}
  \caption{combo+rfi}
\end{subfigure}
\\
\begin{subfigure}[t]{0.33\textwidth}\vskip 0pt
  \imagebox{6cm}{\includegraphics[width=\textwidth]{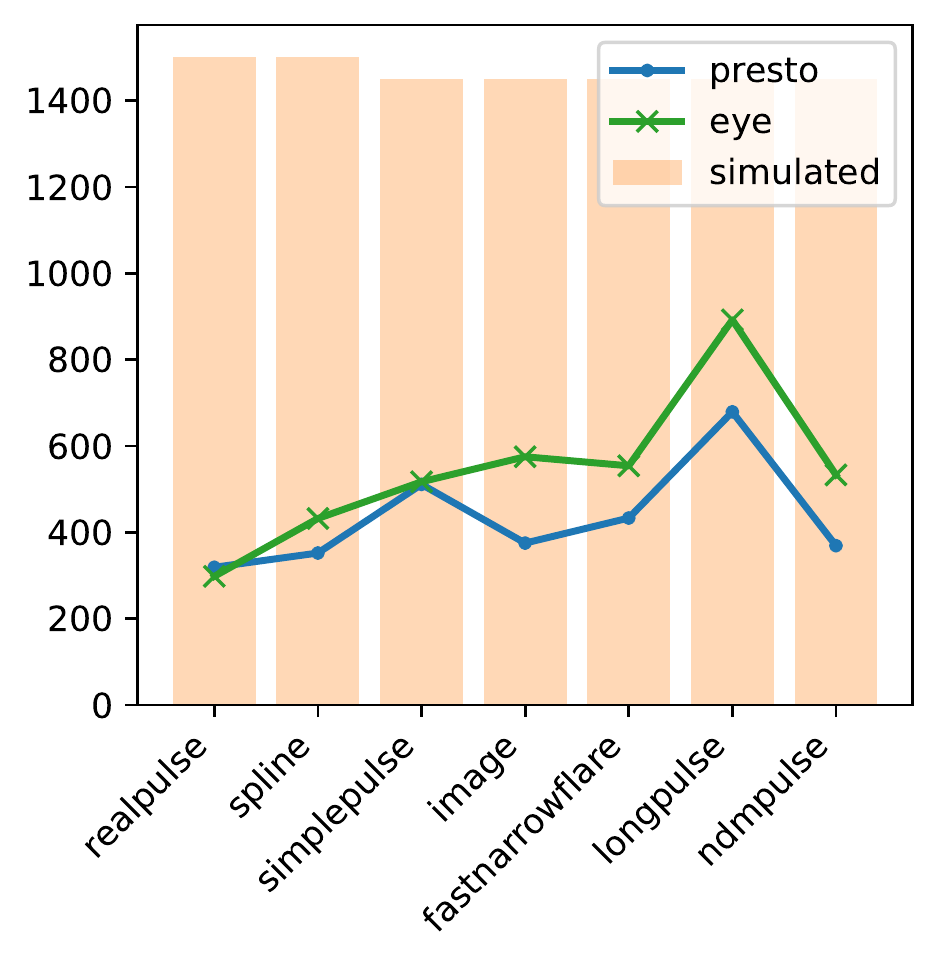}}
  \caption{real+combo}
  \label{fig:tpcount_real+combo_multi_eye}
\end{subfigure}
\caption{True positives detected for the different types of signal for each event groups for multibeam survey using \textsc{PRESTO} and by eye (only for real+combo).}
\label{fig:tpcount_multi}
\end{figure*}

\begin{figure*}
\centering
\begin{subfigure}[t]{0.33\textwidth}\vskip 0pt
  \imagebox{6cm}{\includegraphics[width=\textwidth,height=6cm]{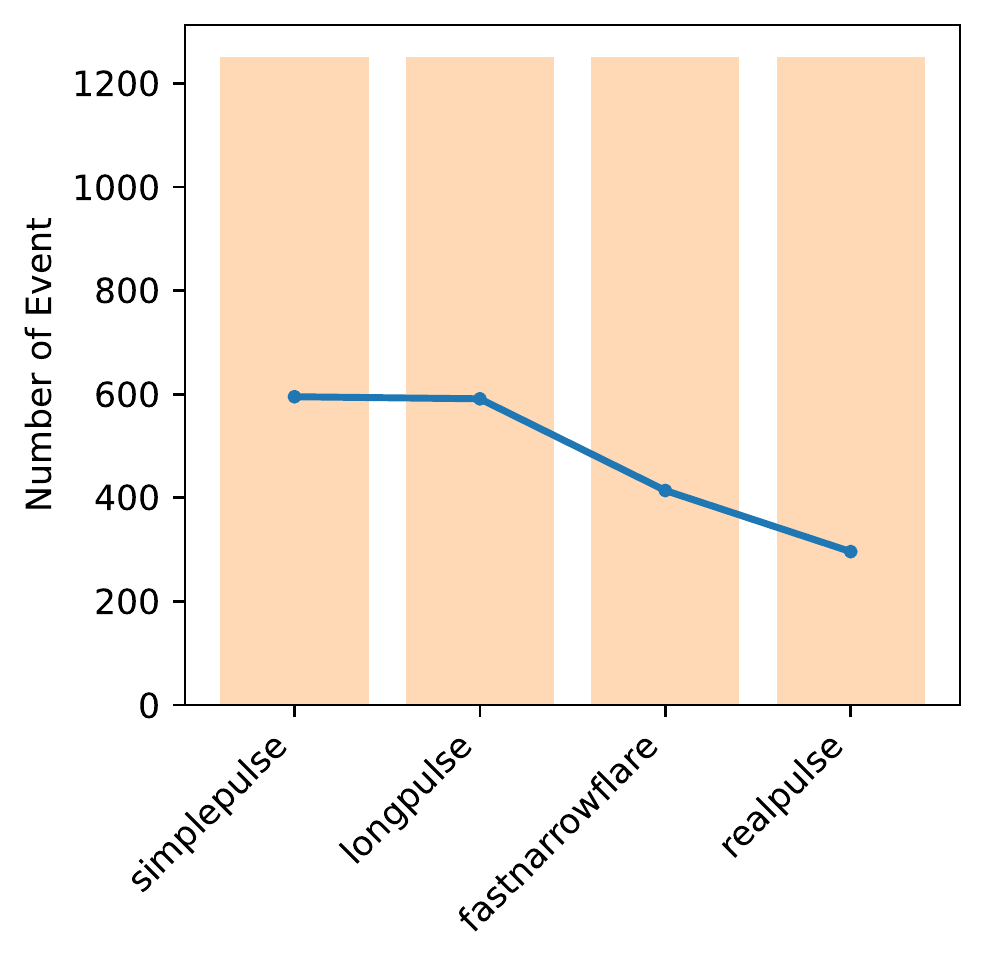}}
  \caption{known+rfi}
\end{subfigure}%
\begin{subfigure}[t]{0.33\textwidth}\vskip 0pt
  \imagebox{6cm}{\includegraphics[width=\textwidth]{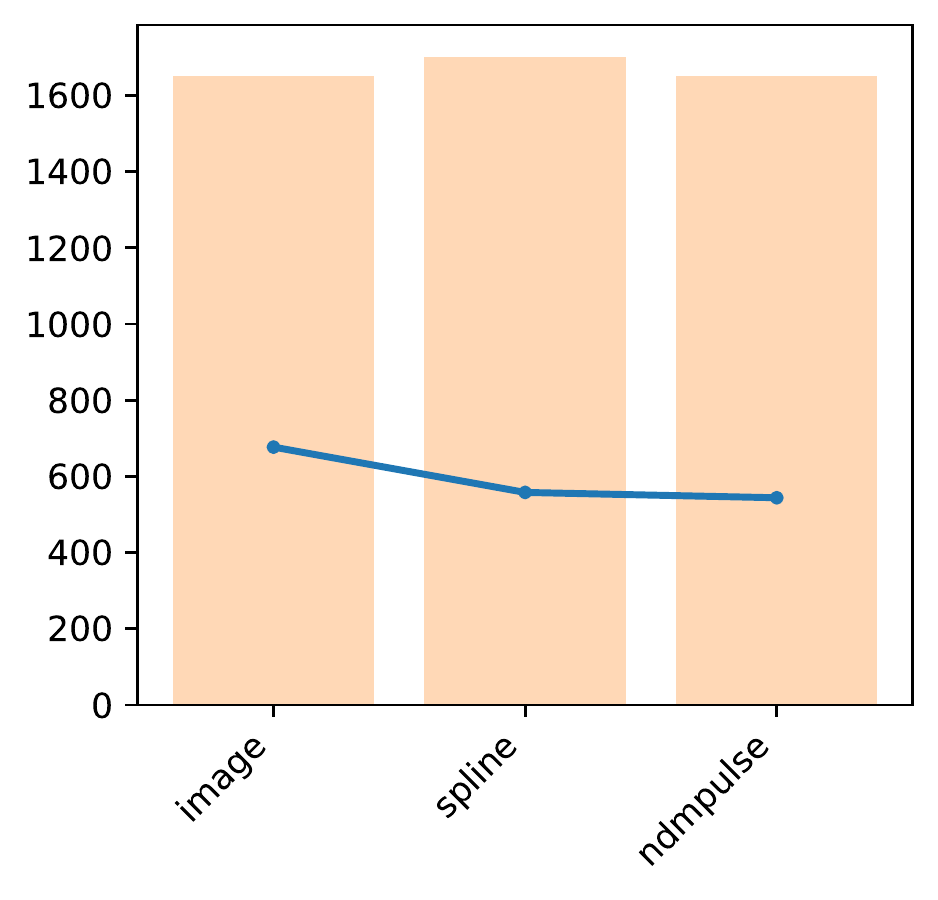}}
  \caption{unknown+rfi}
  \label{fig:tpcount_unknown_paf}
\end{subfigure}%
\begin{subfigure}[t]{0.33\textwidth}\vskip 0pt
  \imagebox{6cm}{\includegraphics[width=\textwidth]{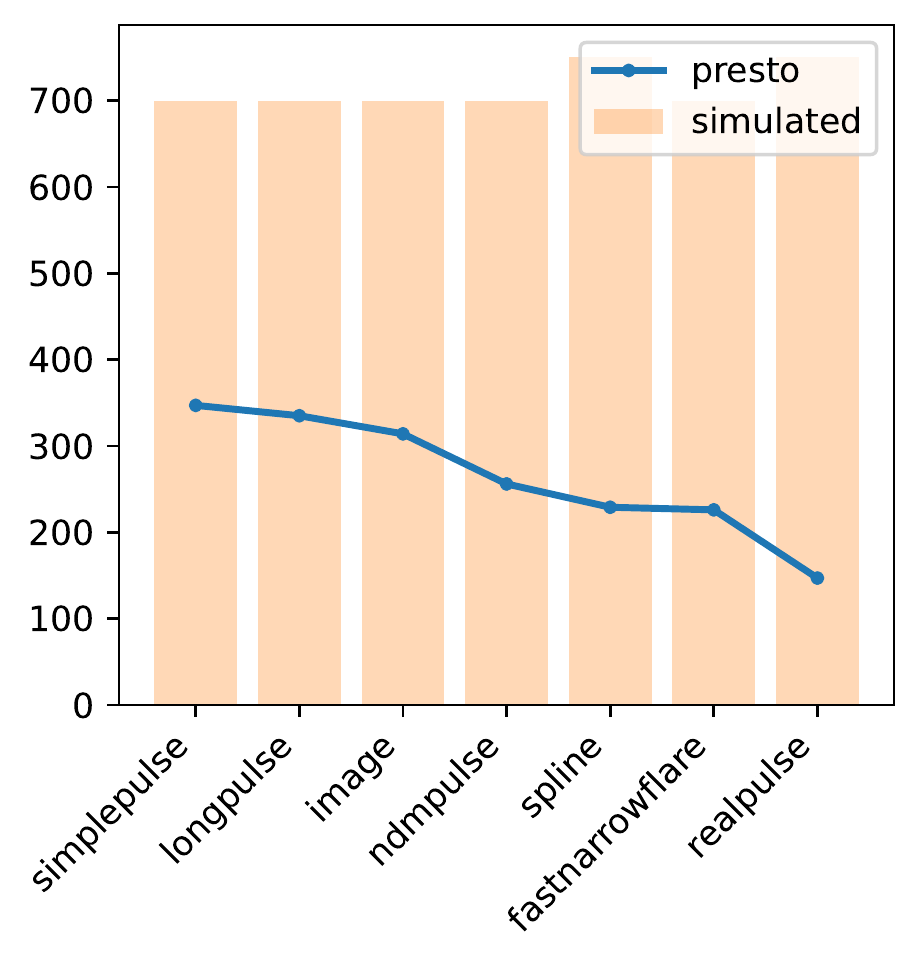}}
  \caption{combo+rfi}
\end{subfigure}%
\caption{True positives detected for the different types of signal for each event groups for PAF survey using \textsc{PRESTO}.}
\label{fig:tpcount_paf}
\end{figure*}

The \texttt{known+rfi} simulations represent idealised single pulses as well as more realistic single pulses whose profiles include frequency dependence, long-period pulse events, and the expected signatures from stellar flares. The \textsc{PRESTO} analysis leads to significantly more false negatives, false positives and fewer true positives than in the simpler cases. The corresponding evaluation metrics for recall, precision and F1 score, all reduce for this simulation, with the fall-out increasing. We group the results by type of simulated event in \cref{fig:tpcount_multi,fig:tpcount_paf} for the multibeam and PAF surveys, respectively. Here we plot the number of simulated and detected events as a function of the source types. The lowest detectability is for real pulses; these are pulses that have frequency structure and do not emit across the entire bandwidth. We therefore expect a reduction in sensitivity to such bursts as \textsc{PRESTO} integrates across the entire bandwidth. This is more noticeable for the PAF survey where the bandwidth is larger than in the multibeam survey.

We note that \textsc{PRESTO} is able to detect 30 to 40 percent of our flare star signatures. The flare star signals contain frequency evolution (again implying less sensitivity with the wider bandwidth of the PAF survey) and contain a linear time-frequency dependence. The linear trend will deviate further from the dispersion curve over the wider bandwidth for the PAF survey. Some of the flare star signals can be modelled relatively well with a positive DM and hence \textsc{PRESTO} does detect such events. However, other events have the equivalent of negative DMs and \textsc{PRESTO} only detects the brightest of these.

\Cref{tab:cmevalmetric} and \cref{fig:tpcount_unknown_multi,fig:tpcount_unknown_paf} show the results for unexpected signals in the simulated data sets. We expect that \textsc{PRESTO} is not an optimal algorithm for such signals, but that it will detect bright signals that can approximately be modelled using a dispersion law. Out of these source types, the injected images are most likely to be detected by \textsc{PRESTO}, whereas the pulse events that do not follow a frequency-squared law or spline curves have similar detectability. 

The combinations (\texttt{combo+rfi} and \texttt{real+combo}) represent the most realistic possible data sets and we discuss these further in \cref{sec:discussion}. 

\section{Discussion} \label{sec:discussion}

\subsection{Using the Data Challenge to Test a New Algorithm} \label{ssec:testnewalgo}

We chose only to use \textsc{PRESTO} in providing our baseline results. There are numerous other algorithms already available for searching for signals in high-time resolution data sets (and other options that can be trialled within \textsc{PRESTO} itself). To compare those algorithms, or newly developed algorithms, with \textsc{PRESTO}, our data challenge can be used. Each frame in our data challenge includes either one injected event, or none. The data collection contains comma-separated values (\texttt{CSV}) files (in the \texttt{gtbaseline\_label} directory) that labels each frame (in the \texttt{label} column) and whether our \textsc{PRESTO} analysis detected an event, or not (in the \texttt{label\_presto} column). Refer to \cref{appsec:sparkesxinfo} for more details.

The results from any algorithm that provides labels for each frame can therefore be compared with the true injections and also with the \textsc{PRESTO} output. The results can be grouped or filtered based on the different properties or types, as illustrated in \cref{fig:truepositiverate_simplepulse,fig:amplitude+width+dm_simplepulse,fig:tpcount_multi,fig:tpcount_paf}. To make the histogram in \cref{fig:truepositiverate_simplepulse}, the parameter values were binned into 20 equal width bins within the minimum and maximum range. To make the plots in \cref{fig:amplitude+width+dm_simplepulse,fig:tpcount_multi,fig:tpcount_paf}, the true positives were grouped by the column type.

As a demonstration, we viewed a subset of the data by eye. For the \texttt{simplepulse} and \texttt{real+combo} datasets for the multibeam survey, we viewed each frame where a signal has been injected, using the \texttt{pfits\_plot} of the \textsc{pfits} package \citep{Hobbs:2021}. We recorded whether, or not, we saw a signal of interest. In \cref{fig:truepositiverate_simplepulse_multi_eye,fig:amplitude+width+dm_simplepulse_multi_eye,fig:tpcount_real+combo_multi_eye}, we compare the methods. The true positive rate is shown for the \textsc{PRESTO} result overlaid on the by-eye analysis in \cref{fig:truepositiverate_simplepulse_multi_eye}. As expected, both analyses detect the majority of the bright events, but the by-eye analysis is less effective than \textsc{PRESTO} at identifying weaker events. Out of the 5000 simulated events for idealised pulses, the FN and FP for the analysis by eye are 3209 and 1791, respectively, which yields the recall evaluation metric of 0.36. As expected, this is worse than the \textsc{PRESTO} recall of 0.53. \textsc{PRESTO} missed 78 FRBs that were detected by eye. The majority of these had high DMs that were above our DM search (see also the left panel in \cref{fig:truepositiverate_simplepulse_multi_eye}). 

The analysis by eye of the injections of all events types into actual observations is summarised in \cref{fig:tpcount_real+combo_multi_eye}. The by-eye analysis in this case is remarkably good and out-performs \textsc{PRESTO} even for the injected idealised FRB-like pulses. This is because the eye easily picks up bright, wide pulses. However, in the presence of RFI, the \texttt{rfifind} procedure cannot distinguish those events from RFI and therefore masks the corresponding parts of the data.

It is relatively easy by-eye to first identify the types of narrow-band RFI present in the real data and then to look for anomalous signals (of any type). The eye is drawn to ``edges'' (for example the boundaries of the images and the spline curves). The eye is not biased towards positive or negative slopes, but was biased towards events occurring near the middle of each frame (short-duration events on the edges were commonly missed).

\subsection{Limitations of the Data Challenge}

The data challenge contains a large amount of data. We have explicitly chosen such data volumes to ensure that the data set can, if required, be divided into training and test data sets (or validation). If, however, the data volume is not sufficient (or the user wishes to produce simulated data sets relevant for their telescope and observing system) then it is relatively easy to produce further data sets using the parameters of the input signals described in this paper. Details on how to produce the simulations using \textsc{simulateSearch} software are available from \citet{Luo+:2022}.

Our data challenge has been divided into frames, which contain either an event or no event. This division leads to three challenges:
\begin{itemize}
  \item Our requirement that high-DM events are constrained to a specific frame implies that all high-DM events start near the beginning of each frame. Any use of this data challenge to train a machine learning algorithm therefore needs to take this into account (potentially by including more than a single frame when training the algorithm).
  \item For very narrow events, an algorithm (including by-eye visualisation) would be more effective by ``zooming-in'' around the event. The choice of how much data is provided to any algorithm therefore may affect the sensitivity of that algorithm.
  \item Real events in actual observations will not be clearly delimited into frames. Any algorithm being developed therefore must work for events that cross sub-integration boundaries.
\end{itemize}

Our datasets contain only relatively short-duration events of $\sim 1$\,second. We are therefore not probing the detectability of longer-duration events through this data challenge.

Even though we have simulated mock multibeam and PAF surveys, we have concentrated on the shape of the features likely in the data streams. We do not claim the completeness of all possible combinations of signals in the dataset. The \textsc{simulateSearch} software has been developed to make it easy to simulate generic signal types. The user can therefore update the data challenge with other types of events as necessary.

\subsection{Towards the Development of New Search Algorithms}

\textsc{PRESTO} is able, as expected, to detect nearly all idealised single pulse events such as single pulses from pulsars and FRBs in conditions of high S/N and low RFI. We therefore do not believe that any new algorithm will be able to improve significantly on \textsc{PRESTO}'s ability to detect such sources (noting that high-DM events will not be found if the searched DM range is not sufficient).

We recommend the development of new algorithms for (1) data sets significantly affected by RFI, both in terms of reducing false positive events and also in detecting actual events in the presence of the RFI, (2) narrow-band FRB events or for pulsars whose scintillation bandwidth is smaller than the observing bandwidth, (3) signatures of flare stars, and (4) unexpected signals that do not follow the dispersion law.

The search for the unknown unknowns is challenging. For bright events (for instance, those that can be seen by eye in the time-frequency dynamic spectra) numerous methods exist, including edge detection algorithms, feature extraction methods, etc. The primary challenge is in identifying a suitable scale for the dynamic spectrum being analysed. If the time-scale being analysed is smaller than the event duration, or the event width then the detectability of the signal will reduce. If the time-scale is much longer than the event duration then the statistics of the measured data will be dominated by the radiometer noise and the event likely to be missed.

Weak ``unknown unknown'' events (those undetectable by eye in dynamic spectra) will be extremely hard to detect unless prior knowledge exists around the possible properties of the signal, or a detailed understanding exists of the underlying noise. We do note that flare star, or similar, events are likely to exist within our archival data sets \citep{Tang+:2022} and therefore recommend searching for linear, as well as quadratic, frequency-dependent signals.

We have chosen to simulate rare events. By definition the data set will therefore have fewer events than ``non-events''. This is also true in the real observations where training data sets for e.g., FRB events are dominated by RFI, noise etc. and only a few actual FRBs are present. It is therefore essential that new algorithms can deal with this imbalance and we explicitly chose to ensure that the injected signals in our data sets remain rare events. A proper handling of the imbalanced classes will likely boost the performance of a classification algorithm. Several methods can be used, for example, resampling the data set to balance the classes when training a supervised machine learning for classification.

We have not explicitly studied how fast \textsc{PRESTO} runs on these data sets, but future algorithms need to be efficient in order to process the large data volumes expected from future surveys. We also note that algorithms that do not rely on multiple de-dispersion steps (e.g., \citealt{Zhang+:2018}, \citealt{Zhang+Wang+:2020}) may be significantly faster than \textsc{PRESTO}.

\section{Conclusion and Future Direction} \label{sec:conclusion}

Here, we have begun to address how current and future telescope high-time resolution surveys can be used to detect signals that are currently unknown. We have presented our SPARKESX dataset of Single-dish PARKES for finding the uneXpected. The SPARKESX dataset includes three mock surveys from the Parkes ``Murriyang'' telescope. 

The SPARKESX dataset is designed to aid in the development of new search algorithms. Event labels exist allowing new algorithms to be compared against our baseline \textsc{PRESTO} results. We envision that this data challenge will useful in the development of advanced statistical, machine learning, and data analysis techniques. The practicality is not limited to astronomy, but also more broadly across multiple disciplines, in particular to the analysis of massive data sets.

Our longer-term plan is to use this data challenge in order to develop both image-processing and machine learning algorithms that can identify anomalies in high-time resolution data streams. We will compare and test these algorithms with the SPARKESX data challenge and then apply to archival Parkes data sets as well as to the future surveys planned using the new cryogenic PAF that will soon be installed at the telescope.

\section*{Acknowledgements}

We thank the anonymous referee for valuable suggestions on the manuscript. We thank Prof.\ Ron Ekers for discussions relating to finding the unknown. We thank Russell Tsuchida for discussions on the dataset. This work was funded by CSIRO's Machine Learning and Artificial Intelligence Future Science Platform. This paper includes archived data obtained through the Parkes Pulsar Data archive on the CSIRO Data Access Portal (\url{https://data.csiro.au/}). The Parkes ``Murriyang'' radio telescope is part of the Australia Telescope National Facility (\url{https://ror.org/05qajvd42}) which is funded by the Australian Government for operation as a National Facility managed by CSIRO. We acknowledge the Wiradjuri people as the traditional owners of the Observatory site.

This research made use of the following software: \textsc{Matplotlib} \citep{Hunter:2007}, \textsc{NumPy} \citep{vanderWalt+:2011,Harris+:2020}, \textsc{pandas} \citep{McKinney:2010,pandas:2020}.

\section*{Data Availability}

The SPARKESX data are available from the CSIRO DAP at \url{https://doi.org/10.25919/fd4f-0g20}. Refer to \cref{appsec:sparkesxinfo} for more details.

\bibliographystyle{mnras}
\bibliography{References}

\appendix
\section{SPARKESX Information} \label{appsec:sparkesxinfo}

The SPARKESX dataset contains two parts:
\begin{enumerate}
\renewcommand{\theenumi}{(\arabic{enumi})}
  \item The main simulation data and injected real data in \texttt{PSRFITS} format files, as described in the paper.
  \item The complementary properties labels of the simulated events for each \texttt{PSRFITS} files in comma-separated values (\texttt{CSV}) format, which are described in \cref{tab:labeldata}.
\end{enumerate}
In addition, the labels for every frames or the ground truth are compiled into a \texttt{CSV} format file for reproducibility of the results. Essentially, it contains all the properties labels that are filled to include non-events as well. The outputs from the baseline method using \textsc{PRESTO} are also appended in the same file. The column information is described in \cref{tab:labeldata}.

The dataset for SPARKESX can be downloaded from the CSIRO DAP at \url{https://doi.org/10.25919/fd4f-0g20}.

\subsection{Folder Layout and Filename Convention}

The layout of the dataset folder is as follows. It contains three directories for the different surveys, namely single-beam from multibeam, 13-beams from multibeam, and PAF. Each directory has 7 for multibeam and 6 for PAF subfolder with the event groups, and 1 for multibeam 13-beam with real+combo subfolder. Within it, the main \texttt{PSRFITS} simulation data and \texttt{CSV} labels are included.

For the \texttt{PSRFITS}, the file naming convention: \texttt{<eventgroup><+rfi>\_<survey>\_<index><\_realfilename>.sf}
\begin{itemize}
  \item \texttt{eventgroup}: Event groups as listed in \cref{tab:simsignalgroup}
  \item \texttt{+rfi}: If simulated RFI is added
  \item \texttt{survey}: Survey name. \texttt{sbeam} for 1-beam of multibeam, \texttt{mbeam} for 13-beam of multibeam, and \texttt{paf} for PAF. For \texttt{mbeam}, additional suffix \texttt{<beam\#>\_t<mbtype>}, where \texttt{beam\#} is the beam number and \texttt{mbtype} is the type of configuration of the receiver (refer to \cref{ssec:injectintoreal}).
  \item \texttt{index}: File number
  \item \texttt{realfilename}: If real observation data is used, the filename of the raw data
\end{itemize}

For the properties label \texttt{CSV}, the file naming convention: \texttt{<eventgroup>\_<survey>\_<index>\_propparam\_label.csv}

In the directory \texttt{gtbaseline\_label}, the files containing the ground truth labels and outputs from \textsc{PRESTO} are stored. Similarly, within it contains directories for the three surveys, single-beam, 13-beam multibeam, and PAF. The file naming convention: \texttt{<eventgroup>\_<survey>\_<index>\_gtbaseline.csv}.

Additionally, in the directory \texttt{image\_irm}, the injected steganography images are provided for reference.

\subsection{File Size}

A summary of the number of files and events is listed in \cref{tab:sparkesxfile}. The file size of each \texttt{PSRFITS} for
\begin{itemize}
  \item 1-beam of multibeam: 49\,MB for all event groups except 101\,MB for real+combo
  \item 13-beam of multibeam: 101\,MB
  \item PAF: 8\,GB
\end{itemize}
The \texttt{CSV} label files are only a few KB each. The \texttt{CSV} ground truth labels and \textsc{PRESTO} outputs files are mainly $\sim 3\,$MB except 6\,MB for real+combo single-beam. The total file size of SPARKESX is 2.34\,TB.

\begin{table*}
  \centering
  \caption{Summary of number of files in SPARKESX dataset.}
  \label{tab:sparkesxfile}
  \begin{tabular}{@{\extracolsep{4pt}}l *{3}{c}@{}}
  \toprule
  Survey & 1-beam of multibeam & 13-beam of multibeam & PAF \\
  \midrule
  Number of frames per simulated file & 1000 & - & 1000 \\
  Number of events per simulated file & 100 & - & 100 \\
  Number of frames per injected file & 2051 & 2051 & - \\
  Number of events per injected file & 205 & 205 & - \\
  Number of event groups (see \cref{tab:simsignalgroup}) & 7 & 1 & 6 \\
  Number of files per event group & 50 & 13 & 50 \\
  Total number of files & 350 & 13 & 300 \\
  \bottomrule
  \end{tabular}
\end{table*}

\subsection{Label Information}

The columns in the \texttt{CSV} label files are separated by a comma (\texttt{,}) delimiter. The label descriptions are listed in \cref{tab:labeldata}.

\begin{table*}
  \centering
  \caption{Label descriptions for simulation and injected SPARKESX data.}
  \label{tab:labeldata}
  \begin{tabularx}{\textwidth}{l @{\extracolsep{\fill}} *{2}{c} X}
  \toprule
  Column Name & Format & Unit & Description \\
  \midrule
  frame & integer & & Frame number, which corresponds to a sub-integration of 1 for the single-beam and 4 for PAF. Not provided for 13-beam multibeam. \\
  iframe & integer & & Index of the frame which is the same as one sub-integration. 1 iframe=1 frame for single-beam and 1 iframe=4 frame for PAF. \\
  type & string & & Type of event: \texttt{simplepulse} for simple pulse, \texttt{realpulse} fpr realistic pulse, \texttt{longpulse} for long period pulse, \texttt{fastnarrowflare} for fast narrow flare, \texttt{slowwideflare} for slow wide flare, \texttt{ndmpulse} for negative DM pulse, \texttt{spline} for spline, \texttt{image} for steganography image \\
  amplitude & float & Jy & Amplitude \\
  log10(amplitude) & float & Jy & Logarithmic base 10 of amplitude \\
  spectralexponent & float & & Spectral exponent \\
  width & float & second & Width \\
  log10(width) & float & second & Logarithmic base 10 of width \\
  dm & float & \pccm & Dispersion measure, DM \\
  fref & float & MHz & Reference frequency \\
  tbw & float & MHz & Total bandwidth for flare \\
  log10(tbw) & float & MHz & Logarithmic base 10 of total bandwidth for flare \\
  driftrate & float & MHz\,s$^{-1}$ & Drift rate for flare \\
  npoint & integer & & Number of points of nodes \\
  arrivaltime & float & second & Arrival time of image \\
  height & float & & Height of image \\
  img\_path & string & & File path of image in the \texttt{image\_irm} directory \\
  filename & string & & Base filename of simulation file \\
  label & integer & & Ground truth label: 0 for no event, 1 for event \\
  label\_presto & integer & & True positive identified by \textsc{PRESTO}: 0 for no event, 1 for event \\
  dm\_presto & float & \pccm & Dispersion measure of event identified by \textsc{PRESTO} \\
  sigma\_presto & float & & Signal-to-noise (S/N) of event identified by \textsc{PRESTO} \\
  time\_presto & float & second & Time of event identified by \textsc{PRESTO} \\
  \bottomrule
  \end{tabularx}
\end{table*}

\subsection{Reading the Data}

The simulated data sets are in \texttt{PSRFITS} format, the same data format as actual high-time resolution survey data sets as recorded by radio telescopes. This implies that any algorithm that has been developed using SPARKESX can directly be applied to peta-bytes of real observations available from e.g., the Parkes pulsar data archive \citep{Hobbs+:2011}. However, this means that new algorithms need to account for the data format and, in particular, the data quantisation to 1 or 2-bits. Users of the data challenge may prefer images. Such data products can be produced once the \texttt{PSRFITS} simulation file is loaded using a package to read \texttt{FITS} file. For example, in \textsc{C}, using the \textsc{CFITSIO}\footnote{\url{https://heasarc.gsfc.nasa.gov/fitsio/}} \citep{Pence:1999,Pence:2010} library, but the user needs to provide their own unpacker. In \textsc{Python}, the \textsc{PulsarDataToolbox}\footnote{\url{https://github.com/hazboun6/PulsarDataToolbox}} can be used using the function \texttt{pdat.psrfits}. Note that this (and other similar tools) only reads bytes from the PSRFITS files. The user then needs to extract individual 1- or 2-bit streams from those bytes. For 1-bit data, this can be carried out using the \textsc{NumPy} \texttt{unpackbits} routine. For 2-bit data a look-up table or more complex routine is needed. Tools such as \textsc{pfits} \citep{Hobbs:2021} and \textsc{dspsr} \citep{vanStraten+Bailes:2011} can also be used to read and process these data files.

\bsp	
\label{lastpage}
\end{document}